\UseRawInputEncoding
\documentclass[a4paper]{article}
\usepackage{blindtext} 
\pdfoutput=1
\usepackage[sc]{mathpazo} 
\usepackage{microtype} 

\usepackage[english]{babel} 

\usepackage[hmarginratio=1:1,top=32mm,columnsep=20pt]{geometry} 
\usepackage[hang, small,labelfont=bf,up,textfont=it,up]{caption} 
\usepackage{booktabs} 

\usepackage{lettrine} 
\usepackage{graphicx}

\usepackage[square,numbers]{natbib}
\usepackage{enumitem} 
\setlist[itemize]{noitemsep} 

\usepackage{abstract} 

\usepackage{titlesec} 
\renewcommand\thesection{\Roman{section}} 
\renewcommand\thesubsection{\roman{subsection}} 
\titleformat{\section}[block]{\large\scshape\centering}{\thesection.}{1em}{} 
\titleformat{\subsection}[block]{\large}{\thesubsection.}{1em}{} 

\usepackage{titling} 

\usepackage{hyperref} 

\usepackage{xcolor}

\begin{document}

\title{Pneumococcus and the stress-gradient hypothesis: a trade-off links $R_0$ and susceptibility to co-colonization across countries}

\author{Ermanda Dekaj $^{1,\dag}$, Erida Gjini $^{1,*}$ \\
\small{$1$ Center for Computational and Stochastic Mathematics, Instituto Superior Tecnico, University of Lisbon, Lisbon, Portugal}\\
\small{ $^\dag $ermanda.dekaj@tecnico.ulisboa.pt,$^*$ Corresponding: erida.gjini@tecnico.ulisboa.pt}}

\maketitle
\begin{abstract}
Modern molecular technologies have revolutionized our understanding of bacterial epidemiology, but reported data across different settings remain under-integrated in common theoretical frameworks. Pneumococcus serotype co-colonization, caused by the polymorphic bacteria Streptococcus pneumoniae, has been increasingly investigated in recent years. While the global genomic diversity and serotype distribution of S. pneumoniae are well-characterized, there is limited information on how co-colonization patterns vary globally, critical for understanding bacterial evolution and dynamics. Gathering a rich dataset of cross-sectional pneumococcal colonization studies in the literature, we quantified patterns of transmission intensity and co-colonization prevalence in children populations across 17 geographic locations. Fitting these data to an SIS model with co-colonization under the assumption of similarity among interacting strains, our analysis reveals strong patterns of negative co-variation between transmission intensity ($R_0$) and susceptibility to co-colonization ($k$). In support of the stress-gradient hypothesis in ecology (SGH), pneumococcus serotypes appear to compete more in high-transmission settings and less in low-transmission settings, a trade-off which ultimately leads to a conserved ratio of single to co-colonization $\mu=1/(R_0-1)k$. Within our mathematical model, such conservation suggests preservation of 'stability-diversity-complexity' regimes in multi-strain coexistence. We find no major study differences in serotype composition, pointing to underlying adaptation of the same set of serotypes across environments. Our work highlights that understanding pneumococcus transmission patterns from global epidemiological data can benefit from simple analytical approaches that account for quasi-neutrality among strains, co-colonization, as well as variable environmental adaptation.
\end{abstract}

\section{Introduction}

One of the most intriguing hypotheses in ecology is the stress-gradient hypothesis (SGH), which predicts that positive interactions in multispecies communities should be more prevalent in stressful environments, while more benign environments should favour competition \citep{bertness1994positive,callaway1997competition,chamberlain2014context}. This hypothesis has been empirically tested, refined and supported mainly by studies of plant systems \citep{pugnaire2001changes,callaway2002positive,eranen2008increasing,maestre2009refining,he2013global}, and only more recently in microbial communities \citep{mccluney2012shifting,fetzer2015extent,hoek2016resource,lawrence2016evolution,piccardi2019toxicity}. For example, \citep{piccardi2019toxicity} studied 4 bacterial species that could degrade toxic industrial fluids and showed that positive interactions in the system were most common at high levels of abiotic stress and intermediate nutrient concentrations where most species could not grow, whereas making the environment more habitable promoted competition. 

In general, how environmental variability (both temporal or spatial) affects species diversity is a highly debated topic in ecology. It has been argued that intermediate disturbance intensity and frequency of disturbances maximize species diversity \citep{connell1978diversity,grime1973competitive}. Changes may come externally (e.g., abiotic resource supplies) or be caused by the organisms themselves, e.g. via resource consumption. Modern coexistence theory argues that such environmental
factors mediate the sign and/or magnitude of interspecific interactions \citep{chesson1994multispecies,hoek2016resource,piccardi2019toxicity}, and ultimately whether species tend to cooperate or compete in a system can critically drive community diversity
and stability \citep{mougi2012diversity,butler2020cooperation}.

Infectious disease epidemiology offers another context where the stress gradient hypothesis can be studied in relation to biodiversity. Infectious diseases often involve multiple pathogen species or multiple strains of the
same pathogen \citep{balmer2011prevalence,griffiths2011nature}. For infectious agents, interactions between pathogens or pathogen strains can take many forms: for example, 
infection by one strain can alter susceptibility to subsequent infection by others, or
simultaneous presence of multiple strains can affect the duration, infectiousness, as well as severity of infection \citep{may1995coinfection,mosquera1998evolution,alizon2013multiple}. While pathogen interactions are recognized to have far-reaching clinical, epidemiological, and eco-evolutionary implications, investigations of the stress-gradient hypothesis in polymorphic epidemiological contexts, to our knowledge, have not been undertaken so far. This may be partly due to the complexity of defining and inferring interactions in multi-strain pathogens, and partly due to the lack of a simple and general enough framework linking "stress" to "interactions" in such systems. 

In this paper, we address this gap, by considering the stress-gradient-hypothesis in an epidemiological multi-strain system, namely \textit{Streptococcus pneumoniae} bacteria, shaped by heterogeneous interactions in co-colonization between multiple serotypes. We study application of SGH using microbial co-colonization prevalence data from different geographical and epidemiological settings and linking them with the predictions of a mathematical model \citep{madec2020predicting,gjini2020key}. \textit{Streptococcus pneumoniae} is an important human pathogen, which despite being carried asymptomatically most of the time in the nasopharynx, can cause acute otitis media, bacterial meningitis, sepsis and pneumonia, in children and vulnerable adults such as the elderly \citep{bogaert2004streptococcus,o2009burden}. The natural habitat of the pneumococci is the mucosa of the human nasopharynx, where a single strain can persist for weeks or months, sometimes in parallel with other strains. The phenomenon of more than one pneumococcal strain colonizing the nasopharynx at the same time, is known as co-colonization, or multiple carriage, which can refer to these being different serotypes, or clones, or distinct strains defined by some trait of interest \citep{brugger2009detection}. 

There are more than 90 serotypes defined by the polysaccharide capsule in \textit{Streptococcus pneumoniae}, which have associations with host immunity and invasive disease \citep{geno2015pneumococcal}, although much genetic diversity exists also within each serotype. Furthermore, there is wide variation in the range of pneumococcal colonization and co-colonization prevelances, as well as serotype and clonal diversity in different parts of the world, and this variability is not completely understood, both when considering pre-vaccine trends and trends in the post-vaccine era. This may reflect variations in transmission environment, study populations with respect to age, ethnicity, underlying host genetics, socioeconomic living conditions, vaccine coverage, and differences in sampling, isolation techniques and transport, and methods of sampling \citep{libwea2020prevalence,rivera2009multiplex,ercibengoa2012dynamics}. The health conditions of study subjects at the time of sampling may also vary and impact results. In addition, overcrowding and poor hygiene in some settings may favour the interchange of infecting microorganisms. Finally, different geographical settings offer also different climate conditions, including temperature, humidity, light exposure, and eco-epidemiological contexts in terms of other co-circulating pathogens and diseases which can interfere with pneumococcus transmission. 

How to interpret all this variation? While a single model for natural carriage cannot take all these factors into account in individual detail, an abstract and sufficiently general epidemiological framework can encapsulate all these variations into a global transmission intensity parameter for the focal colonizing species of interest, in this case, pneumococcus basic reproduction number $R_0$. On the other hand, the complexities of inter-strain interactions and local composition at the level of serotypes and co-circulating clones can be encapsulated into a global parameter of susceptibility to co-colonization $k$.

A recent mathematical model proposed by \citep{madec2020predicting}, helps to formalize and integrate this approximation, by describing the processes of single and co-colonization in a polymorphic microbial ecosystem by an SIS (susceptible-infected-susceptible) epidemiological model with $N$ similar strains and co-infection in two timescales (a \textit{fast} neutral one, and a \textit{slow} non-neutral timescale). The strains exhibit slight variation in pairwise co-colonization susceptibility coefficients $K_{ij}=k+\epsilon \alpha_{ij}, (0\leq\epsilon\ll 1)$, enough to drive hierarchical epidemiological dynamics between them, in a broad spectrum of stable and unstable possible configurations. In the matrix, $K_{ij}\geq 1$ describe pairwise facilitation between $i$ and $j$, while entries below 1, denote pairwise competition or antagonism between $i$ and $j$. Such variability in interactions between strains, creates frequency-dependent competition in epidemiological transmission and leads to explicit replicator dynamics over a slow timescale between $N$ strains \citep{madec2020predicting}. 

\begin{figure*}[!ht]
\centering
\includegraphics[width=\linewidth]{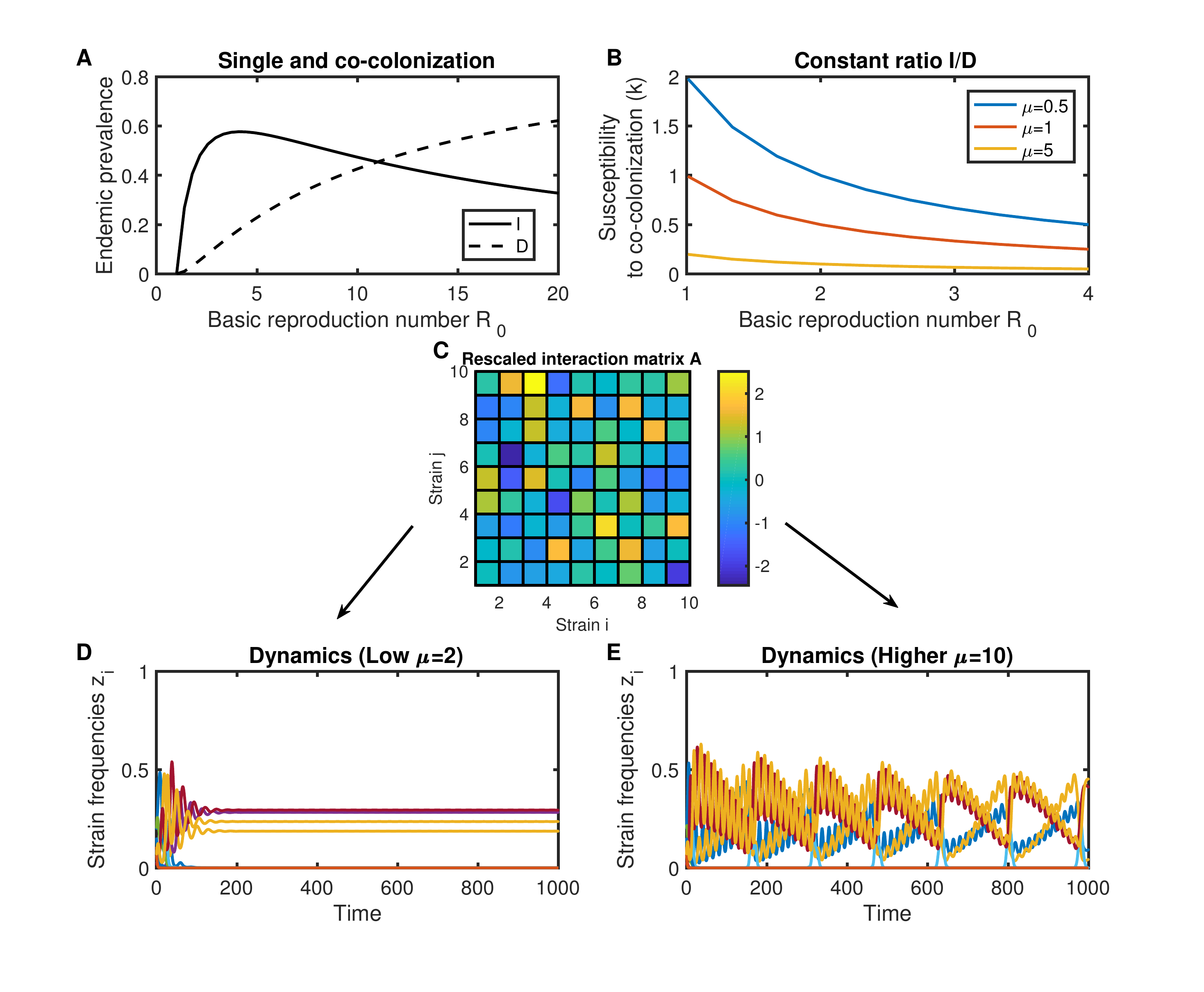}
\caption{\textbf{Theoretical features of the SIS co-colonization model with $N$ similar strains by \citep{madec2020predicting}}. \textbf{A.} Theoretical variation of single and co-colonization prevalence at endemic equilibrium as a function of transmission intensity, captured by the global basic reproduction number $R_0$ (the co-colonization susceptibility parameter $k$ is fixed). \textbf{B.} Expectation of co-variation between $R_0$ and mean susceptibility to co-colonization, $k$, that preserves the same ratio of single to co-colonization in the endemic system $\mu=I/D$. \textbf{C.} In a quasi-neutral SIS co-colonization system with multiple strains, pairwise susceptibility coefficients $K_{ij}$ can vary relative to a reference $k$: $K_{ij}=k+\epsilon \alpha_{ij}$ and $\epsilon$ small ($0 < \epsilon \ll 1$). The matrix $A=(\alpha)_{ij}$ captures such relative variation in interaction strengths between `colonizer' and `co-colonizer' entities, illustrated here for a normal distribution with mean 0 and standard deviation 1 and $N=10$ strains (as in \citet{gjini2020key}). \textbf{D-E.} Keeping the same relative variation strengths $A$, but changing the global $\mu$ (e.g. from low to high value) in the system, gives rise to a very different qualitative dynamics between strains \citep{gjini2020key}. In this case, we varied $k$, keeping $\beta=3, R_0=2$ constant. The multi-strain frequency dynamics, over the slow timescale $\tau=\epsilon t$, are given by the replicator system: $\frac{d}{d\tau} z_i = {\Theta z_i \cdot[ \sum_{j\neq i} \lambda_i^j z_j -\mathop{\sum}_{1\leq k<j\leq N} (\lambda_j^k+\lambda_k^j) z_jz_k ]},\quad i=1,\cdots,N$ where $\lambda_i^j=\alpha_{ji}-\alpha_{jj} -\mu(\alpha_{ij}-\alpha_{ji})$ denote pairwise invasion fitnesses between any two strains, and $\Theta=\beta\left(1-\frac{1}{R_0}\right)\left( \frac{\mu}{2(\mu+1)^2-\mu}\right)$ gives the speed of the dynamics. Any variation in $\mu=\frac{1}{k(R_0-1)}$ can be obtained either by changing $k$ or changing $R_0$. Evidently, for constant relative interaction strengths $A$, to preserve the same qualitative dynamics (i.e. pairwise invasion fitness matrix $\Lambda$) under such mean-field shifts in one parameter, an opposite change in the other parameter is necessary (see also Figure \ref{fig:dynSGH} for an illustration, Supplementary Text S2 and \citep{gjini2020key} for more mathematical details).}
\label{fig:model}
\end{figure*}

Studying mathematically the feedbacks between neutral and non-neutral dynamics, and the range of system behavior for different mean-field parameters and special cases (see Figure \ref{fig:model}), they have suggested that the single-to-cocolonization ratio $\mu$ may act as a gradient for dynamic complexity of multi-strain coexistence \citep{gjini2020key}. In the limit of low $\mu\to 0$, for the same relative variation in co-colonization coefficients between strains ($\alpha_{ij})$, the system tends to (multi-) stable fixed point coexistence and a smaller number of strains coexisting per site. In contrast, in the limit of large $\mu \to \infty$, the system displays oscillatory coexistence behavior of more strains, including limit and heteroclinic cycles, corresponding to well-studied classical examples of the replicator equation in evolutionary game theory \citep{Rank2008,Allesina2011}. 

As the single to co-colonization ratio $\mu$ in their model \citep{madec2020predicting} is given by $\frac{1}{(R_0-1)k}$, i.e. it encodes explicitly a trade-off between transmission intensity and mean interaction parameter in co-colonization, they argue that keeping $\mu$ constant under different environmental pressures and conditions would support application of the stress-gradient hypothesis \citep{gjini2020key}, where in `harsher' conditions (lower carriage prevalence, $R_0$) $k$ would show an increase, and conversely, in more `benign and favourable' conditions (higher carriage prevalence, $R_0$) $k$ would be lower. Host susceptibility to co-colonization once colonized by a pneumococcus serotype has so far been estimated in the range 10-50\% \citep{auranen2010between,numminen2013estimating,lipsitch2012estimating,gjini2016direct}. If any trade-off variation in $k$ in response to $R_0$ is verified, it should imply upregulation/downregulation of cooperative interactions between strains in susceptibility to co-colonization if the same set of strains is found to occur across systems. In contrast, if different strains are associated to different settings, the trade-off should be verified as a result of differential selection, e.g. entities with different competitive abilities being effectively selected in different $R_0$ environments. 

It is precisely this phenomenon that we investigate here, inspired by the $N$-strain co-colonization model of \citep{madec2020predicting,gjini2020key}. We use published pneumococcal studies in the literature, and comparatively analyze their reported data, to offer a fresh quantitative exploration of the stress-gradient hypothesis in epidemiology.

\begin{table*}[!ht]
\caption{\label{data} \textbf{Summary of the pneumococcus cross-sectional data}. Here, we summarize all epidemiological data extracted from the literature: country, year when the study was conducted, sample size, mean age of children, when reported, otherwise its range, susceptibles' prevalence, single colonization prevalence, double colonization prevalence, the number of serotypes reported, and the article reference for each survey. Locations are ordered in increasing order of transmission intensity. Two studies from Greece \citep{syrogiannopoulos2002antimicrobial} and Venezuela \citep{rivera2011carriage}, were excluded from analysis after an outlier test was performed by regressing $n_I/n_D$ against $n_s^{-1}$ and keeping only those data points whose residuals contained 0.} 

\begin{tabular}{p{0.1cm}p{1.9cm}p{1.6cm}p{0.8cm}p{0.6cm}p{1.8cm}p{1.8cm}p{1.8cm}p{1cm}p{1cm}}
\hline 
Id. & Location & Year  & Sample size $n$ & Age (y) & Susceptible $n_s (S)$ & Single-col. $n_i (I)$ &  Co-col. $n_d (D)$ & Nr. serotypes $N$ & Ref. \\ \hline
1&Iran&2009&1291&$<2$&850 (65.8\%)&213 (16.5\%)&228 (17.7\%)  & 30  & \citep{tabatabaei2014multiplex} \\
2&Spain&2004-2005&105&$<3$&11 (10.5\%)&53 (50.5\%)&41 (39\%) & 21+NT  & \citep{ercibengoa2012dynamics} \\
3&Venezuela-1&2004-2005&50&$<5$&15 (30\%)&25 (50\%)&10 (20\%) & 7 & \citep{rivera2009multiplex}\\ 
4&Iceland-1&2009&514&$<6$&123 (23.9\%)&299 (58.2\%)&92 (17.9\%) & 12+NT & \citep{hjalmarsdottir2016cocolonization} \\ 
5&Netherlands&2009&1169 &$<2$&366 (31.3\%)&623 (53.3\%)&180 (15.4\%) & 32+NT &  \citep{wyllie2016molecular}\\ 
6&Bangladesh&2012&212&$<5$&87 (41\%)&98 (46.2\%)&27 (12.7\%)&  27+NT & \citep{saha2015detection}\\ 
7&Nepal&2010-2013&600&$<2$&303 (50.5\%)&237 (39.5\%)&60 (10\%)& 65+NT & \citep{kandasamy2015multi}\\ 
8&Portugal-1&2001&270&$<6$&97 (35.9\%)&143 (53\%)&30 (11.1\%)  & 43+NT  & \citep{valente2012decrease};\citep{sa2009changes}\\
9&Norway-2&2008&602&$<5$&119 (19.8\%)&411 (68.3\%)&72 (12\%) &  34+NT & \citep{vestrheim2010impact} \\
10&Portugal-2&2006&449&$<6$&130 (29\%)&276 (61.5\%)&43 (9.6\%) & 43+NT & \citep{valente2012decrease};\citep{sa2009changes} \\
11&Norway-1&2006&611&$<5$&136 (22.3\%)&415 (67.9\%)&60 (9.8\%)  &  32+NT & \citep{vestrheim2010impact} \\
12&India&2004-2005&510&$<5$&232 (45.5\%)&249 (48.8\%)&29 (5.7\%) & 34+NT  & \citep{sutcliffe2019nasopharyngeal}\\
13&Switzerland&2015-2016&287&$<7$&139 (48.4\%)&134 (46.7\%)&14 (4.9\%) & 10 & \citep{brugger2009detection}\\
14&Brazil&2010&242&$<6$&101 (41.7\%)&128 (52.9\%)&13 (5.4\%)& 25 &  \citep{rodrigues2017pneumococcal}\\
15&Denmark&1999-2000&437&$<6$&190 (43.5\%)&225 (51.5\%)&22 (5\%) & 35 & \citep{harboe2012pneumococcal}\\
16&Iceland-2&2008&514&$<6$&143 (27.8\%)&341 (66.3\%)&30 (5.8\%) & 12+NT  &  \citep{hjalmarsdottir2016cocolonization}\\
17&Vietnam&2010-2011&350&$<5$&210 (60\%)&130 (37.1\%)&10 (2.9\%) & 14+NT  &  \citep{dhoubhadel2014bacterial} \\
18&Mozambique& 2009-2010&927&$<5$&144 (15.5\%)&739 (79.7\%)&44 (4.7\%) & 44+NT  & \citep{adebanjo2018pneumococcal} \\
19&Ghana&2011&848&$<6$&574 (67.7\%)&260 (30.7\%)&14 (1.7\%) & 39+NT  & \citep{dayie2013penicillin}\\
20&Kenya&2009-2010&1087&$<5$&104 (9.6\%)&939 (86.4\%)&44 (4\%)&  56+NT & \citep{kobayashi2017pneumococcal}\\
\hline
\textit{Outliers} &  &   &   &  &  & & &  & \\
\hline
 & Greece & 1997-1999 & 2448  & $<2$ & 1682 (69\%)  & 751 (30.7\%)&15 (0.6\%)  & & \citep{syrogiannopoulos2002antimicrobial} \\
 &  Venezuela-2 & 2006-2008 & 1004 & $<2$ & 733 (73\%)  & 266 (26.5\%)&5 (0.5\%) & & \citep{rivera2011carriage}\\
\hline
\end{tabular}
\end{table*}


\section{Methods}

\subsection{Data collection and preparation}

We have reviewed published papers reporting nasopharyngeal carriage of \textit{S. pneumoniae}, including co-colonization, in different countries. In the literature, we selected cross-sectional studies conducted in young children, typically $<7$ years old, where they reported pneumococcal prevalence of colonization (\% of carriers), prevalence of co-colonization (\% of carriers), and predominantly sought studies performed on healthy children, although this was not used as a strict criterion. For example, we found one study in Switzerland involving children with respiratory infection \citep{brugger2009detection}, and this was also included, based on the assumption that carriage is a pre-requisite for disease, and hence overall patterns of carriage in diseased and healthy subjects are expected to be similar. Data were extracted from these studies for the following variables: prevalence of \textit{S. pneumoniae} carriage, co-colonization prevalence, country, first author, year the study was conducted, age group of participants, detection method for \textit{S. pneumoniae}, study period before or after vaccination, season or months of the year when the samples were collected, reported pneumococcus serotypes, and information on the gender of participants, when available. 

\begin{figure*}[!h]
\centering
\includegraphics[width=\linewidth]{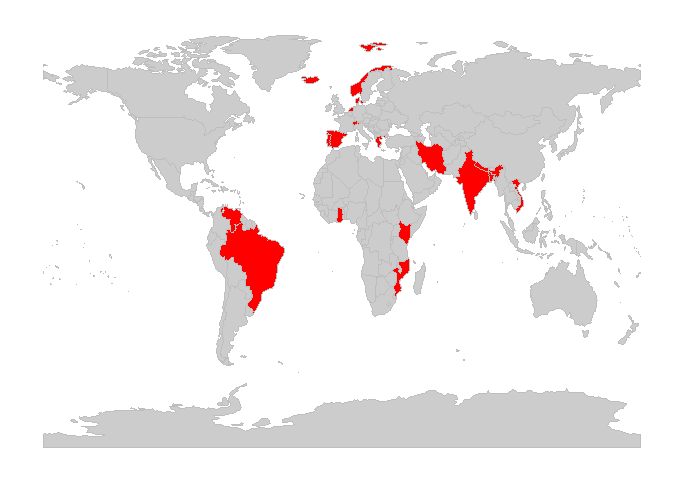}
\caption{\textbf{Country map.} Map of the geographical locations from around the world, where the  cross-sectional studies reported single- and co-colonization prevalences of \textit{Streptococcus pneumoniae} carriage in predominantly healthy children (Table \ref{data}).}
\label{fig:map}
\end{figure*}

After retrieving a large number of pneumococcus articles and epidemiological surveys, we found 22 articles eligible (see Table \ref{data}). The reason for exclusion of studies was either because there was no co-colonization information declared, or the age group did not correspond to our target age. Also, we intentionally selected only cross-sectional studies to match with the mathematical model assumptions, and this condition made us exclude 4 articles that reported co-colonization rates from a longitudinal design (Finland \citep{syrjanen2001nasopharyngeal}, Peru \citep{nelson2018dynamics}, South Africa \citep{manenzhe2020characterization}, Gambia \citep{chaguza2021carriage}). The selected studies were conducted in different countries from four continents: Africa, Asia, Europe and South America. See Figure \ref{fig:map} for a visualization of the countries where the co-colonization reports were found.
We included three instances of studies in the same countries: Iceland-1 and Iceland-2 \citep{hjalmarsdottir2016cocolonization}, where two different methods were used for quantifying prevalence of multiple carriage, Portugal-1 and Portugal-2, conducted before and after implementation of the PCV7 vaccine \citep{valente2012decrease}, and Norway-1 and Norway-2 \citep{vestrheim2010impact}, also conducted in the pre- and post-vaccine era. Most of the studies corresponded to surveys conducted in children in the pre-vaccine era.
We checked for outliers in our data. As outliers resulted to be the studies conducted in Greece \citep{syrogiannopoulos2002antimicrobial} and Venezuela-2 \citep{rivera2011carriage}. These two studies studies appeared to have extremely low values for cocolonization prevalence $<1\%$, very distinct from the rest of the dataset. We removed these studies from our analysis leading to 20 articles in the final set of cross-sectional studies used. Finally we analyzed these studies under a common framework. The data were entered in an Excel spreadsheet for further analysis (See Supplementary files). 

In Table \ref{data} we provide a summary of the multi-country data used in our analysis. In our dataset, the prevalence of susceptible children $S$ in the total sample of the survey, was varying from about 10\% \citep{kobayashi2017pneumococcal} to 70\% \citep{dayie2013penicillin}, the prevalence of single colonization, $I$, was varying from about 16\% \citep{tabatabaei2014multiplex} in Iran to 86\% \citep{kobayashi2017pneumococcal} in Kenya, and the prevalence of co-colonization, $D$, was varying between 1.7\% in Ghana \citep{dayie2013penicillin} and 39\% in Spain \citep{ercibengoa2012dynamics}. Sample sizes across studies vary between 50 and 1291 children in Venezuela \citep{rivera2009multiplex} and Iran \citep{tabatabaei2014multiplex} respectively. The most common \textit{Streptococcus pneumoniae} serotypes shared among \textit{all} studies, demonstrating a high rate of occurrence globally, were: "14"  "19F" "6A"  "6B", a majority of which are included in PCV7 (4, 6B, 9V, 14, 18C, 19F and 23F).

\begin{figure*}[h!]
\centering
\includegraphics[width=\linewidth]{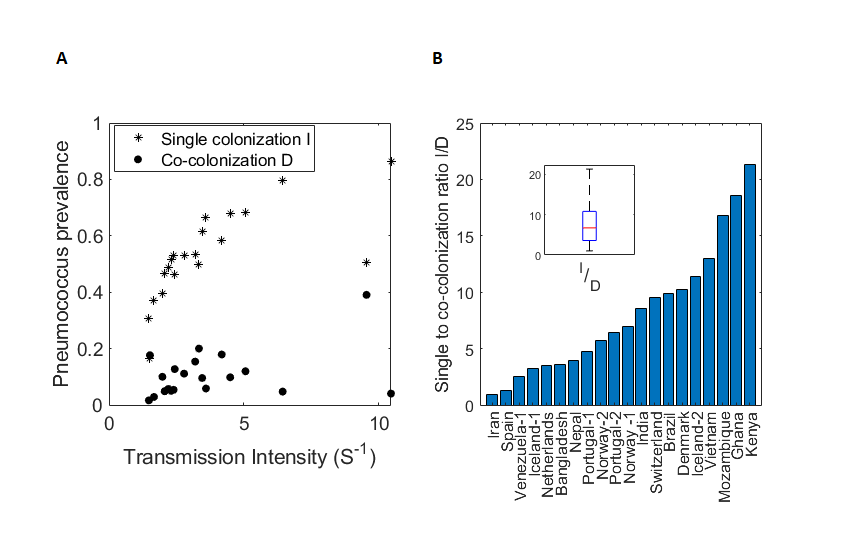}
\caption{\textbf{Visualization of collected epidemiological data.} \textbf{A.} Plot of single colonization and co-colonization prevalences as a function of inverse susceptibles prevalences, reported in the cross-sectional studies of pneumococcus carriage (Table \ref{data}). \textbf{B.} Empirical single-to co-colonization ratios for all countries plotted in increasing order from lowest value reported in Iran, to highest value reported in Kenya.}
\label{fig:datacol}
\end{figure*}

\begin{figure*}[h!]
\centering
\includegraphics[width=\linewidth]{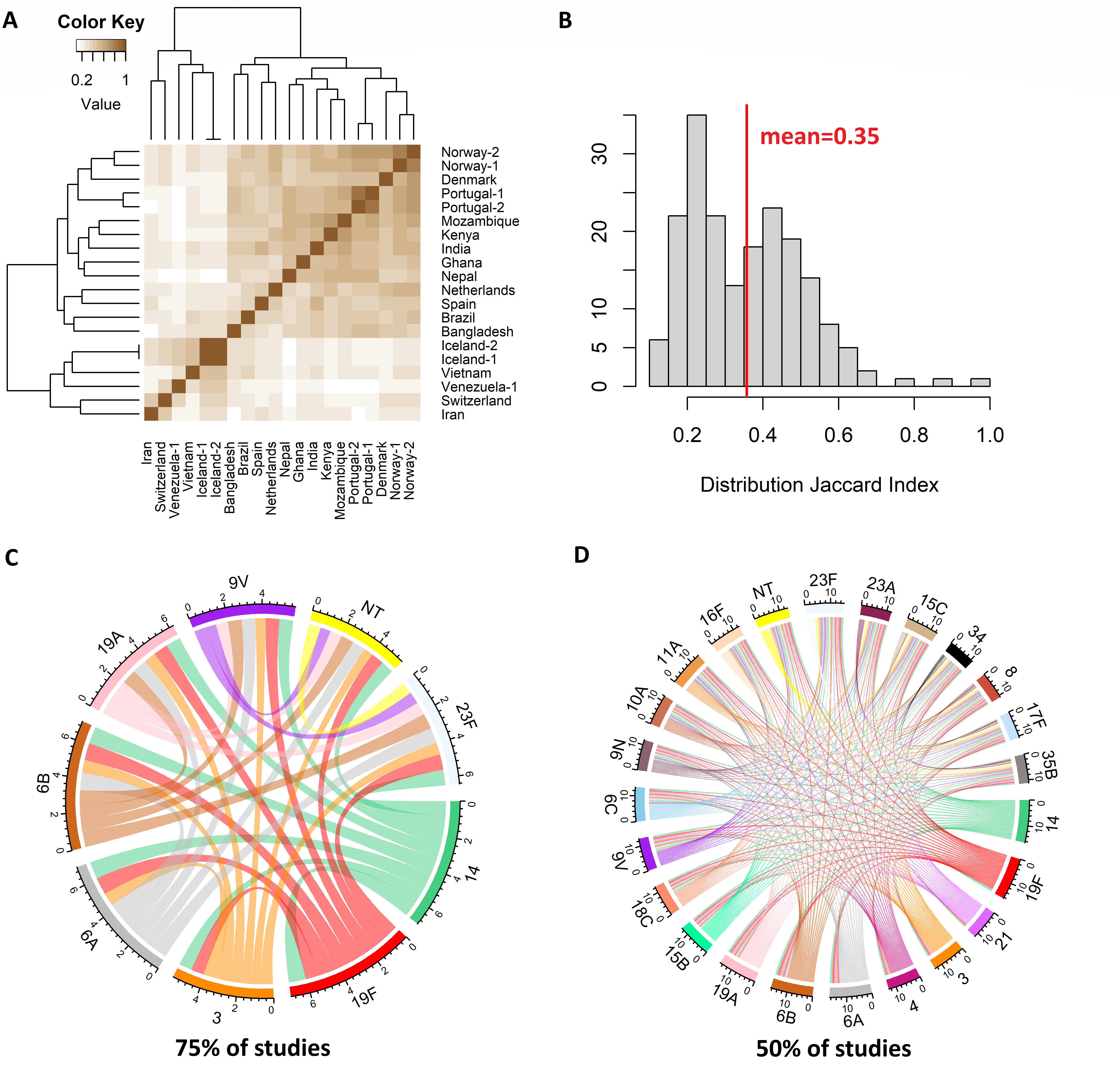}
\caption{\textbf{Serotype composition data across studies.} \textbf{A.} A heatmap of pairwise Jaccard indices across studies with superimposed hierarchical clustering based on this index similarity. A high Jaccard index for two studies denotes a relatively high number of shared serotypes vs. unique serotypes in each study. \textbf{B.} The distribution of Jaccard indices, where the red vertical line denotes the arithmetic mean. \textbf{C.} Co-occurrence network for serotypes found in more than 75\% of studies. \textbf{D.} Co-occurrence network for serotypes found in more than 50\% of studies. In Figures \ref{fig:serotypes}-\ref{fig:bignetwork}, we report more details on the serotype composition and frequencies found in the studies. To visualize the co-occurrences, we used the chord diagram from library circlize in R with function chordDiagram() (\citep{circlize}), where the network represents connections between serotypes (entities). Each serotype is represented by a fragment on the outer part of the circular layout, proportional to its total number of occurrences in the study set. Then, arcs are drawn between each pair of entities. The size of the arc is proportional to the number of times two serotypes were found together in the same study. }
\label{fig:JaccardNetwork}
\end{figure*}

\subsection{The SIS co-colonization model framework}

Following the SIS model in \citep{madec2020predicting}, we assume that the susceptible and colonized host prevalences $S$ and $T$ have already reached their steady-state values $S^{\ast}$ and $T^{\ast}$ (\textit{fast} dynamics), in each country, at the time of the survey, and that the basic reproduction number $R_0$ is higher than 1. Thus, in each study, there is an endemic equilibrium for global prevalence of pneumococcal carriage:

\begin{equation}\label{ST}{(S^{\ast},T^{\ast}) } = (\frac{1}{R_0}, 1 - \frac{1}{R_0}) \end{equation}

Furthermore, when considering the subdivision of carriage into single and co-colonization, the equilibrium values for the proportion of children in these compartments are given by: 
\begin{equation}\label{ID} I^{\ast} =\frac{R_0-1}{R_0[1+(R_0-1)k]} \quad D^{\ast}= \frac{k(R_0-1)^2}{R_0[1+(R_0-1)k]},\end{equation}
where the ratio of single to co-colonization at endemic equilibrium is $$\mu =\frac{I}{D} = \frac{1}{(R_0-1)k}.$$  We will drop the asterisks from now onward, assuming always an equilibrium in these global quantities. Notice that the equilibrium assumption is often adopted in mathematical models of colonization to estimate epidemiological parameters be it from longitudinal data, manifested as constant time-inhomogeneous hazards \citep{auranen2010between,lipsitch2012estimating,numminen2013estimating} or from cross-sectional prevalence data \citep{gjini2017geographic}.

The explicit analytical formulae above derived from the model \citep{madec2020predicting}, under the endemic equilibrium assumption, imply that using the different datasets, we can link the reported proportions of carriers of a single serotype and the proportion of double carriers, to the analytic expressions for single infection prevalence $I$, and double infection prevalence $D$, from this model. Thus, we can obtain the basic reproduction number $R_0$ and mean interaction coefficient in co-colonization $k$ from exactly solving two equations for two unknowns.

Using the SIS model \citep{madec2020predicting} for epidemiological dynamics of \textit{Streptococcus pneumoniae} serotypes (see Supplementary material S2), under the assumption of quasi-neutrality, we can estimate site-specific basic reproduction number as:
\begin{equation}\label{eq:R0}{ \hat{R_0} } = \frac{1}{1 - T}, \end{equation}
and the mean coefficient of susceptibility to co-colonization between serotypes as:  
\begin{equation}\label{eq:k} {\hat{k}} =\frac{D}{I}\frac{1} {\frac{1}{1 - T} -1}, \end{equation} using the data proportions of hosts in different compartments ($T, I, D$), as shown in Table \ref{data}. 

\subsection{Serotype analysis and occurrence in reported studies} %
Each study reports the serotypes they detected by different methods like: multiplex PCR, Quelling method, culture, conventional. We retrieved all serotypes these studies reported, and we collected the information in an Excel Worksheet, noting that some studies report all serotypes they detected meanwhile other studies summarised only some of the serotypes. For example, in the Bangladesh study \citep{saha2015detection}, the article mentions they found a total number of 46 serotypes, but we managed to retrieve from the actual data in their publication only 27 serotypes, besides the non-typable. On average the numbers of serotypes reported per study varied between 7 and 56 (Table \ref{data}). The total number of serotypes appearing at least once in our dataset (20 epidemiological settings) is 102, where commonly identified serotypes include: "6A"  "6B"  "14"   "19F". Then, we analyzed the occurrence and co-occurrence these serotypes had in all the studies of our dataset. We analyzed the number of studies each serotype was reported and the number of times each serotype was co-encountered with any of the other serotypes present in different studies. The co-occurrence patterns between serotypes are summarized in Figure \ref{fig:JaccardNetwork}C-D (See also Figures \ref{fig:serotypes}-\ref{fig:bignetwork} for more details). For details on pneumococcus diversity in terms of serogroup composition of the global  dataset see also Figures \ref{fig:serogr1}-\ref{fig:serogr3}.)
\subsection{Similarity indices between studies}

\paragraph{Epidemiological parameters ($R_0,k$)}
 After estimating site-specific basic reproduction number $R_0$ and co-colonization susceptibility $k$, we also calculate the epidemiological parameter distance ($d_{ij}$) between any two studies as the Euclidean distance between their corresponding ($R_0$, $k$) points in a 2-d space, as $d_{ij}$ = $\sqrt{(R_0^i-R_0^j)^2+(k^i-k^j)^2}$.
 
 \paragraph{Single to co-colonization ratio ($\mu$)}
 We considered also pairwise distances in the single-to-colonization ratio reported in any two studies, as an additional epidemiological indicator of transmission similarity and constraints. 

\paragraph{Serotype composition}
After extracting the identities of the serotypes reported in all cross-sectional epidemiological studies, to compare any two sites on the basis of their serotype composition, we used the Jaccard index. This was calculated as an intersection index between the serotypes reported in study $i$ and $j$ divided by the union of all the serotypes in both studies. The higher this index is, the higher the proportion of common serotypes between the two studies, and the lower the index is, the lower the proportion of serotypes common between studies. Thus, for each pair of studies we have: $$J_{ij} = \frac{|G_i\cap G_j|}{|G_i\cup G_j|},$$ where $G_i$ and $G_j$ are the serotypes reported in studies $i$ and $j$. Non-typeable serotypes (NT) were considered as one separate indistinguishable entity (NT equivalent to an extra 'serotype').

\paragraph{Geographic distances} Geographic distances between any two study sites were evaluated, country-wise, using dataset {dist\_cepii} from the library cepiigeodist in R. The distances are calculated following the great circle formula, which uses latitudes and longitudes
of the most populated city or of its official capital.

\paragraph{Clustering based on similarity}
Based on the similarity measures above, we typically also performed data clustering to examine whether distinct groups of studies would emerge. For the Jaccard indices and geographic distances, we considered the default hierarchical clustering emerging from the \textit{heatmap} function in R. With regards to single to cocolonization ratios ($\mu$), we used k-means clustering functions in R. 

\section{Results}

\subsection{Inferred epidemiological parameters for each setting}
From applying the model to prevalence data of single and double colonization across all the locations, we are able to find the maximum-likelihood values of the basic reproduction number, $R_0$ and the reference interaction parameter in co-colonization, $k$, as the exact solution of two mathematical equations (Eqs. \ref{eq:R0}-\ref{eq:k}), under the assumption of an endemic epidemiological equilibrium of the SIS model. Analyses are performed using R software (version 4.0.3). See Table \ref{estimates} for results of this estimation for all countries. 

\begin{table*}[t!]
\caption{\label{estimates} Estimated epidemiological parameters for each Streptococcus pneumoniae cross-sectional carriage study in young children, across 17 geographical locations, using the SIS model \citep{madec2020predicting}.} 
\begin{tabular}{p{2cm}p{3.5cm}p{3cm}p{4cm}}

 \hline 
\textit{Location}  & \textit{Single to cocolonization ratio} & \textit{Basic reproduction number} & \textit{Mean susceptibility to co-colonization} \\
&  $\hat{\mu}$ (95\% CI)  & $\hat{R_0}$ (95\% CI) & $\hat{k}$ (95\% CI) \\  \hline

Iran &0.93 (0.66-1.06)&1.52 (1.5-1.58)&2.06 (1.9-2.6) \\
Spain& 1.29 (0.27-2.27)&9.55 (7.5-21)&0.09 (0.07-0.18) \\
Venezuela-1  &2.5 (0.51-4.3)&3.33 (2.94-5.56)&0.17 (0.12-0.43) \\
Iceland-1  &3.25 (2.03-3.86)&4.18 (4.02-4.87)&0.1 (0.09-0.13) \\
Netherlands &3.46 (2.45-3.88)&3.19 (3.1-3.5)&0.13 (0.12-0.16) \\
Bangladesh &3.63 (1.62-4.69)&2.44 (2.33-2.9)&0.19 (0.16-0.32) \\
Nepal &3.95 (2.49-4.71)&1.98 (1.92-2.14)&0.26 (0.23-0.35) \\
Portugal-1 &4.77 (2.33-6.23)&2.78 (2.62-3.31)&0.12 (0.1-0.19) \\
Norway-2 &5.71 (3.31-7.03)&5.06 (4.78-6.11)&0.04 (0.04-0.06) \\
Portugal-2 &6.42 (3.56-7.8)&3.45 (3.3-4.01)&0.06 (0.06-0.09) \\
Norway -1 &6.92 (4.12-8.25)&4.49 (4.27-5.27)&0.04 (0.04-0.06) \\
India &8.59 (4.95-10.85)&2.2 (2.13-2.43)&0.1 (0.08-0.14) \\
Switzerland   &9.57 (4.37-12.86)&2.06 (1.98-2.35)&0.1 (0.08-0.17) \\
Brazil  &9.85 (4.09-13.93)&2.4 (2.28-2.86)&0.07 (0.06-0.13) \\
Denmark &10.23 (5.44-13.16)&2.3 (2.22-2.56)&0.08 (0.06-0.12) \\
Iceland-2 &11.37 (6.31-14.33)&3.59 (3.43-4.15)&0.03 (0.03-0.05) \\
Vietnam& 13 (5.98-18.24)&1.67 (1.62-1.8)&0.12 (0.09-0.21) \\
Mozambique&16.8 (10.05-20.38)&6.44 (6.14-7.69)&0.01 (0.009-0.015) \\
Ghana&18.57 (10.35-24.34)&1.48 (1.45-1.55)&0.11 (0.09-0.18) \\
Kenya&21.34 (12.64-26.2)&10.45 (9.79-12.35)&0.00496 (0.004-0.007) \\
\hline
\textit{Global mean}   & 8.1 (4.38-10.4) & 3.73 (3.47-4.9) &  0.19 (0.17-0.28) \\
\textit{Global mode}  & 3.55 (2.47-4.7) &  2.42 (2.3-2.4)  &  0.097 (0.08-0.18) \\
[1ex]

 \hline
\end{tabular}
\end{table*}

From these data, the highest basic reproduction number $R_0$ emerges in the study in Kenya ($R_0=10.45$), whereas the lowest transmission intensity are displayed in the studies in Ghana and Iran ($R_0 \approx 1.5$) in Iran. The highest susceptibility to co-colonization $k$ emerges in Iran ($k=2.06$), and the lowest in the study in Kenya ($k=0.005$). Notice that with the exception of Iran, all the other coefficients support competition between pneumococcus serotypes in co-colonization ($k<1$), indicating that a pre-existing serotype in colonization makes it harder for a second serotype to co-colonize the same host. In contrast, Iran's value for $k$ above 1 indicates average facilitation between colonizer and co-colonizer serotypes in co-colonization in this epidemiological setting.

To obtain confidence intervals for these parameters, we then accounted for sample size and variability in sample sizes across countries. We calculated 95\% confidence intervals for each estimated $R_0$ and $k$, using the multinomial distribution. In our case, we adopted an empirical simulation procedure (500 simulations), assuming a multinomial distribution for the three variables: number of susceptible hosts $n_S$, number of single infected $n_I$, number of coinfected $n_D$, taking into account the sample size of each study ($n=n_s+n_I+n_D$) and the maximum-likelihood 3-dimensional probability vector for $p=[S,I,D]$, with $S+I+D=1$. For each new realization of ($n_S, n_I, n_D$), using the equations (\ref{eq:k}-\ref{eq:k}), we obtain new estimates for $R_0$ and $k$, consistent with that sample size and the reported prevalences in that location. Finally, for each study we report the 95\% confidence intervals for all such estimates of $R_0$ and $k$, from the limits of the 2.5\% and 97.5\% lower and upper quantiles, capturing uncertainty due to sampling process (Table \ref{estimates}). From this simulation procedure (besides distributions for $R_0$ and $k$, as illustrated in detail in Figures \ref{fig:doubleerrorbar}-\ref{fig:hists}), we also obtained 95\% uncertainty intervals for the single-to-co-colonization ratio in each study (Figure \ref{fig:empiricalmu_err}).


\subsection{Testing for the stress-gradient hypothesis across settings}
After collecting all the data from different countries, and fitting the SIS epidemiological model to infer site specific values for $R_0$ and $k$ \citep{madec2020predicting}, we tested for the Stress Gradient Hypothesis, as predicted by this model \citep{gjini2020key}. In order for SGH to hold, the requirement within our model is that the ratio of single-to-cocolonization does not co-vary with $R_0$, i.e. is constant, as $R_0$ varies, thus implying a trade-off between $R_0$ and $k$ values. In our dataset the ratio of single to co-colonization varies from 0.93 to 21.34 in all studies (see Fig.\ref{fig:datacol}), but there is no significant trend for its co-variation with $R_0$ (linear regression for slope different from zero, p-value>0.1).

The expectation for $\mu$ constant is that if the basic reproduction number $R_0$ increases in a setting, the susceptibility to co-colonization should decrease in that same setting. This implies in settings with high prevalence of carriage, pneumococcus serotypes should display higher competition, whereas in settings of lower $R_0$, $k$ should increase, hence increasing more cooperation and facilitative interactions in co-colonization between serotypes. 
When considering the relationship $\mu=\frac{1}{(R_0-1)k}$ in logarithmic scale, the model becomes linear, and hence we used linear regression between the estimated $log(k)$ and $log(R_0-1)$ in each study as a test for the stress-gradient hypothesis, expecting a slope of -1 (Model: $y=a+bx$), expecting:
$$\log(k)=-\log(\mu)-\log(R_0-1).$$

\begin{figure*}[!ht]
\centering 
\includegraphics[width=0.8\linewidth]{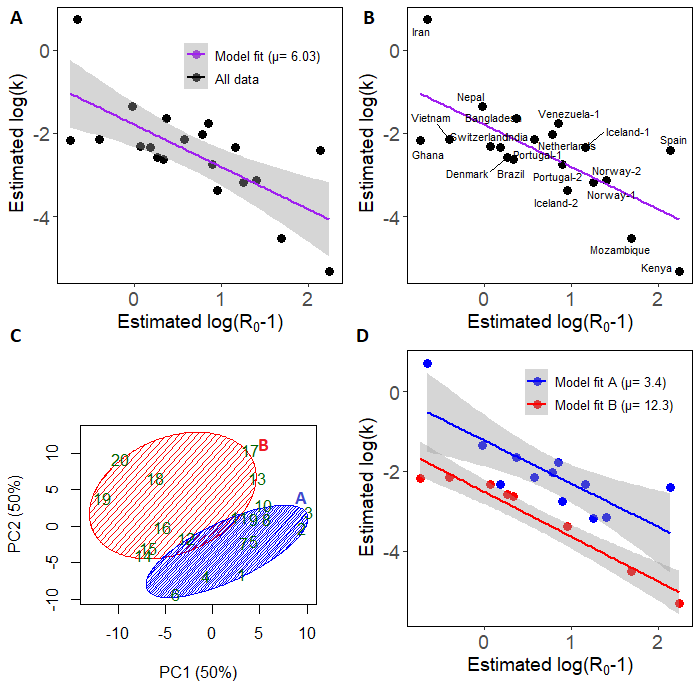}
\caption{\textbf{Testing for the Stress-Gradient-Hypothesis in pneumococcus.} \textbf{A.} Fitting the model to $(\hat{R_0},\hat{k})$ combinations across countries. In this figure, we see the model (linear function) in logarithmic scale. Thick purple line denotes the estimated best-fitting line with $\mu_{SGH}=6.03$. The shading delimits the lower and the upper limits for the 95\% confidence interval estimated by the model for $\mu$ [3.56, 10.21]. (Multiple R-squared:  0.5024,	Adjusted R-squared:  0.4747). \textbf{B.} Best-fitting regression line with slope -1.0133 (from estimated $\mu_{SGH}$= 6.03 from the model), with country names super-imposed for completeness. The negative slope is significant with p-value=0.0005. From the model SGH, we expect the fit line to be with slope -1. \textbf{C.} Emergence of two clusters of studies according to empirical $\mu$ values (high and low ratio of single-to-cocolonization). Visualization in terms of a bivariate plot visualizing a partition (clustering) of the data. All observations are represented by points in the plot, using principal components or multidimensional scaling. We estimated 2 clusters, by computing k-means clustering. Next, the wss (within sum of square) is drawn according to the number of clusters. The location of a bend (knee) in the plot is generally considered as an indicator of the appropriate number of clusters. Here, around each cluster an ellipse is drawn. Blue colour is the indicator for the group A (12 studies), and red colour is the indicator for the group B (8 studies). \textbf{D.} Applying linear regression to each group of studies separately. The slopes of the regression are preserved, both very close to 1 (as expected from the model), together with their significance, while from the intercepts we estimate $\mu_{SGH}^A=3.4$ and $\mu_{SGH}^B=12.3$, very close to the groups' geometric means of empirical $I/D$ ratios.}
\label{muclust}
\end{figure*}

This linear regression confirmed our hypothesis (Fig.\ref{muclust}A-B), for we obtained as best-fitting parameters an intercept equal to -1.796 (p-value = 1.14e-06 ) and a slope equal to -1.0133, very close to 1 (p-value= 0.000468). This implies the best-fitting single to-co-colonization ratio that matches our data is $\mu_{SGH}=6.03$ with 95\% confidence interval [3.56, 10.21].  The mean-squared-error for the model-data differences in estimated $k$, (MSE) for this $\mu$ is equal to 0.15, and coefficient of determination for the model adj. $R^2$=0.4747. 

Notice that this best-fitting line has a distinct predicted $\mu$ from the value obtained using simply the empirical arithmetic mean of $I/D$ values of all countries (equal to 8.1 $\pm 5.79$). The mean squared error of the model-predicted $k$ corresponding to this value ($E[I/D]=8.1$) is 0.17, higher than the error for $\mu_{SGH}=6.03$ (see Figure \ref{fig:crudeoptim}). On the other hand, the $\mu_{SGH}$ resulting from the model is very close to the geometric mean of the empirical ratios of single to co-colonization in our dataset (Geo.mean [I/D]=6.08).

\subsection{Emergence of two epidemiological settings: lower and higher ratio $I/D$}
The variability in empirical single-to-colonization ratios $\mu$, reported across studies (Figure \ref{fig:datacol}), led us to explore similarity and clustering based on this epidemiological quantity. We performed study clustering according to empirical $I/D$ values. Our analysis attempted to find studies that were below and above the global mean, and revealed two clusters of studies, as shown in Figure \ref{muclust}C: \begin{itemize}
    \item \underline{Low-$\mu$ settings:} "Iran" "Spain" "Venezuela-1" "Iceland-1"   "Netherlands" "Bangladesh" "Nepal"       "Portugal-1"  "Norway-2"    "Portugal-2"  "Norway-1"    "India" - grouped in blue color, with arithmetic mean of empirical $I/D$ ratios equal to 4.28, i.e. co-colonization about 4 times less frequent than single colonization.
    \item \underline{High-$\mu$ settings}: "Switzerland" "Brazil"      "Denmark"     "Iceland-2"   "Vietnam"     "Mozambique" "Ghana"       "Kenya", -grouped in red color, with arithmetic mean of empirical $I/D$ ratios, about 3 times higher, equal to 13.84, implying co-colonization of children by more than one serotype is much less likely, about 14 times less frequent than single colonization.
\end{itemize} 

Next, we applied the linear regression for the stress-gradient hypothesis, to each group separately in logarithmic scale (Fig. \ref{muclust}D). We obtained significant slopes very close to -1 in both cases. For group A (blue line), we obtained slope=-1.082 [95\%CI: -0.45 -1.72] (p-value= 0.00351) and intercept=-1.225 (p-value=0.00182), confirming the trade-off between $R_0$ and $k$ around a lower constant $\mu_{SGH}^A = 3.4$ [95\%CI: 1.78 - 6.52]. For the linear model applied to the second group, B (red line), we obtained a slope = -1.123 [95\%CI: -0.84, -1.4](p-value= 6.41e-05) and intercept= -2.5159, (p-value=1.02e-06), maintaining again the expectation from the SGH for this set of studies, but now corresponding to a higher $\mu_{SGH}^B$ = 12.38 [95\%CI: 9.09-16.85]. These model-estimates for $\mu_{SGH}$ are close to the empirical geometric means of empirical values of single to co-colonization ratios in the two groups (Geo.mean $[I/D]^A=3.62$, and Geo.mean $[I/D]^B=13.25$). 

Finally, we verified in a single global regression model that this separation of data in two groups statistically improves model fit, including cluster-id as another variable, (Model: lm($\log(k) \sim \log(R_0-1)\times$ Clustering)). The model with two clusters, preserves the same slope, while significantly improving the fit (p-value=6.691e-06, Multiple $R^2= 0.8037$,	Adj. $R^2=  0.7669$), compared to the model with just a single global $\mu_{SGH}$ (Multiple $R^2=  0.5024$,	Adj. $R^2=  0.4747$). This indicates the potential relevance of two epidemiological environments, where pneumococcus serotype coexistence can be found, for explaining the global distribution of $R_0$ and $k$: in low and high single to co-colonization ratios (although finer-scale resolution is also plausible as shown in Figures \ref{fig:optimalclusters}- \ref{fig:3clustSGH}, but requiring more studies for high statistical power).

\begin{figure*}[!ht]
\centering
\includegraphics[width=\linewidth]{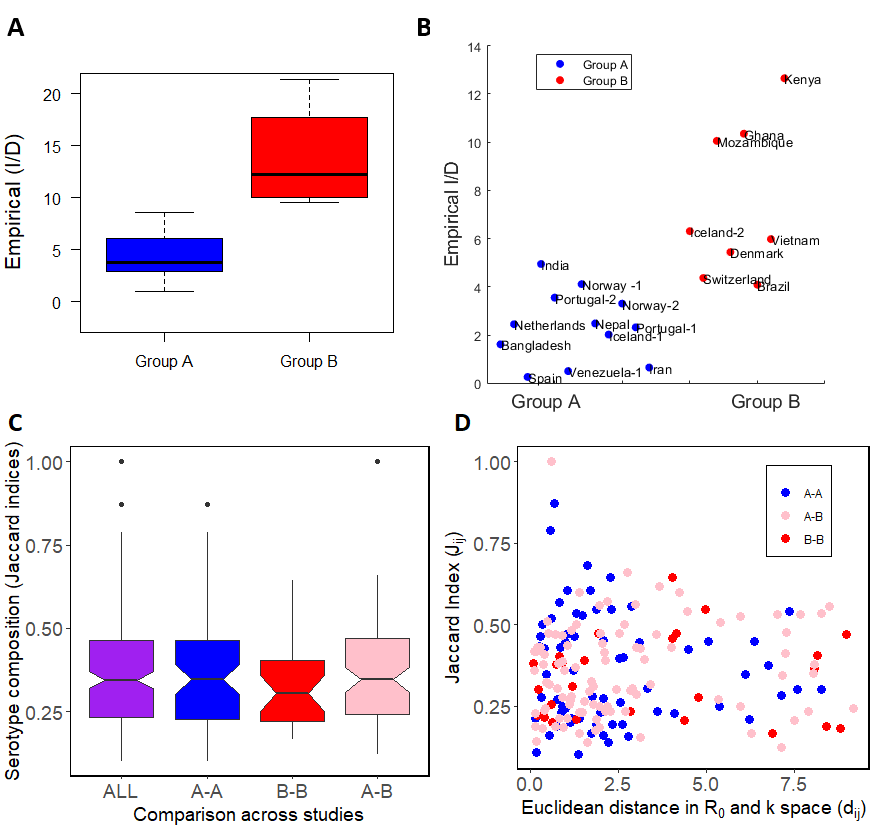}
\caption{\textbf{Comparing serotype composition similarity among studies.} \textbf{A.} The boxplot of empirical single to cocolonization ratios in the identified subsets of studies, group A and group B. \textbf{B.} Specification of studies belonging to the low $\mu$ ($\mu^A_{SGH}=3.4$) and high $\mu$ ($\mu^B_{SGH}=12.38$) group, respectively 12 and 8 studies.  \textbf{C.} Jaccard indices within the data set. Each boxplot refers to the distribution values of Jaccard indices: for all pairs of studies ALL; for pairs of studies within the low-$\mu$ group: A-A, within the high-$\mu$ group: B-B, and comparisons across groups: A-B. Neither grouping has significant differences in Jaccard index means (p-value is greater than 0.05).  \textbf{D.} We illustrate the full co-variation with a scatter plot of Euclidean distance in epidemiological parameters between any two studies ($d_{ij}$) and pairwise Jaccard Index ($J_{ij}$), for the relevant comparisons. There is no significant linear relationship between serotype composition similarity and epidemiological similarity between sites (linear regression, p-value >0.5) .}
\label{Jaccarddheboxplot} 
\end{figure*}

\subsection{Serotype composition similarity across studies}
\paragraph{Jaccard indices and global similarity}
To investigate whether there were any differences between serotype compositions reported in all the studies across different geographical locations, and $(R_0,k$) settings, we first examined the Jaccard indices and compared them across the entire dataset, as well as with respect to the two groups: low and high $\mu$. 
In Figure \ref{Jaccarddheboxplot} we see the comparison of empirical single to co-colonization ratios across the two groups (\ref{Jaccarddheboxplot}A), a summary of the regression results (\ref{Jaccarddheboxplot}B), and how the Jaccard index values vary with respect to these groupings (\ref{Jaccarddheboxplot}C-D). In Fig.\ref{Jaccarddheboxplot}C, t-test results confirmed that serotype compositions do not vary significantly across groups. In particular, we did not find that for pairs of studies in group (A-A comparisons in $J_{ij})$) or in group B (B-B Jaccard indices) or between groups, the serotype compositions are significantly different, (Two-sample Welch t-test, p-value>0.1). We examined also how serotype similarity indices varied continuously along the epidemiological gradient $(R_0,k)$ as described by the Euclidean distance between site pairs (Fig. \ref{Jaccarddheboxplot}D) . We did not find a significant linear relationship, neither for all studies, neither when focusing on subgroup comparisons, i.e. studies in group A or group B only. 

This suggests that, when looked at this resolution level, even though sites may differ in epidemiological parameters, while still satisfying the trade-off between carriage prevalence (transmission intensity $R_0$) and competition in co-colonization ($k$), it is ultimately the same serotypes which broadly occur everywhere, hence pointing to the same serotypes adapting in their mutual interaction coefficients in response to $R_0$ alteration. 

\begin{figure*}[!ht]
\centering
\includegraphics[width=0.9\linewidth]{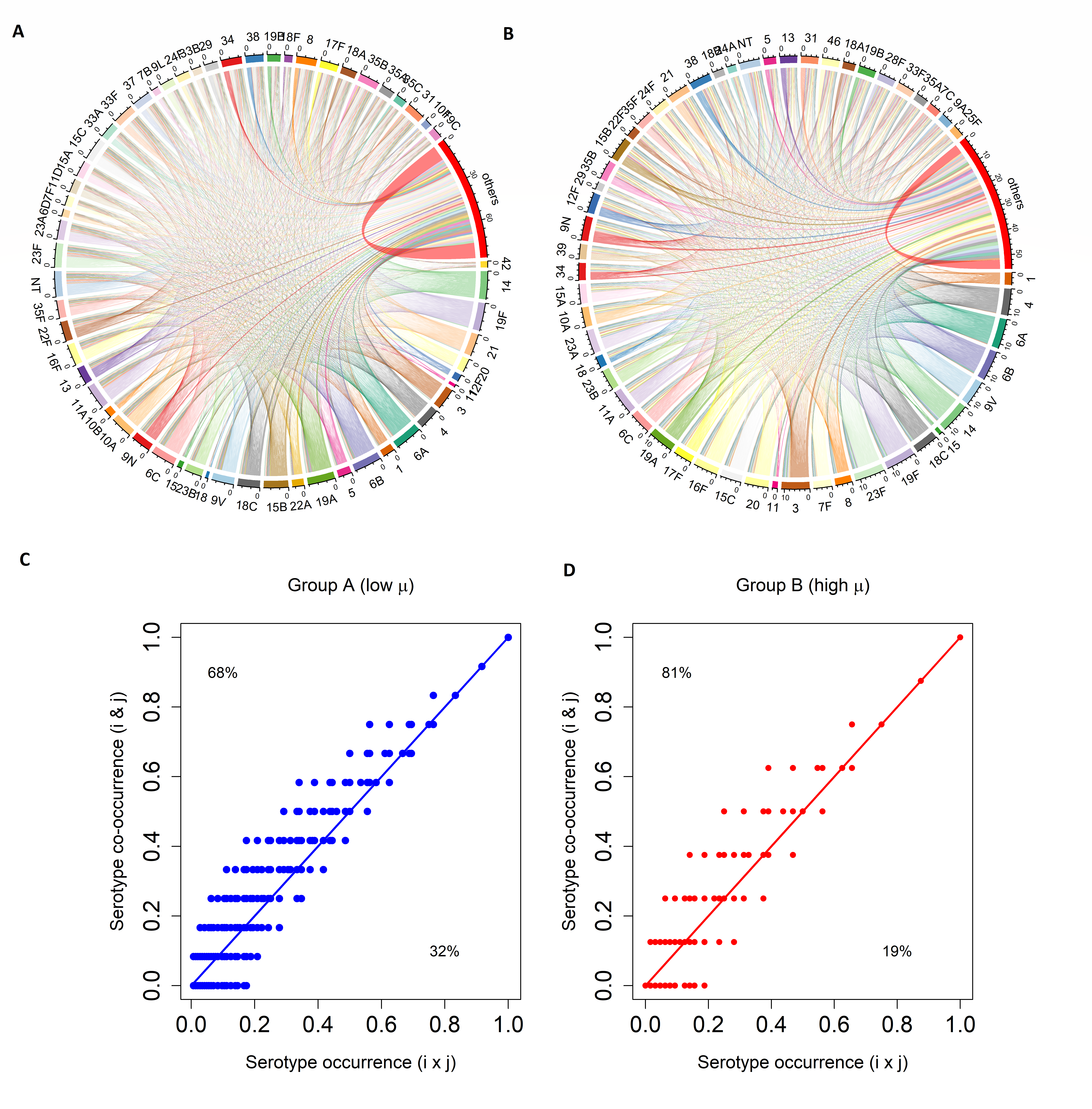} 
\caption{ \textbf{Serotype co-occurrence network for the low- and high-$\mu$ groups of studies.} Here we visualize all serotypes found in reported studies separated in two subgroups according to similarity in the ratio of single-to-colonization ($\mu$ clustering). \textbf{A.} Group A studies are 12 studies (low $\mu=3.4$). \textbf{B.} Group B studies are 8 studies (high $\mu=12.3$ ). Together studies in group A report 92 serotypes in total, and those in group B report 80 serotypes in total. In the network, we denote by `others', serotypes that are found in just 1 study. The thickness of each edge connecting two serotypes varies with the frequency of their co-occurrence in the same study/geographical setting. See also Figures \ref{fig:rankfreq++} and \ref{fig:propser++}. \textbf{C-D} Serotype co-occurrence patterns in each group according to chance and independence. Here we show the
scatter plot of probability of co-occurrence of two serotypes in the same site, vs. the probability of reported serotype $S_i$ times the probability of reported serotype $S_j$ in the global dataset. In group A (low $\mu$ studies), the percentage of serotype pairs co-occurring below chance is 32\%, meanwhile in group B this percentage is lower, 19\%. Fisher's Exact Test for the odds ratio yields p-value $<$ 2.2e-16 with $OR=2.00$ and 95 percent confidence interval: [1.79, 2.24], confirming an association between $\mu$ and serotype pairs co-occurrence patterns, with relatively higher odds of serotype co-occurrence above chance in settings of higher ratio $I/D$ (here group B).}
\label{fig:2networksaccgrAgrB}
\end{figure*}

\paragraph{Serotype identities and occurrence patterns} Then we examined more closely the actual serotype compositions, in terms of precise identities, and their number of occurrences within each subset of the data (A and B) hence going deeper into any potential subtler differences between the two subgroups obtained from $\mu$ clustering. In group A, we found 5 unique serogroups: 45, 48, 32, 40, 27, which can be said to be associated with the epidemiological settings of low ratio of single to co-colonization. In group B, we found 3 unique serogroups: 46, 44, 47, which may be associated with epidemiological settings of high ratio of single to co-colonization. In any case, the group-specific serotype and serogroup occurrences were manifested in rare serotypes/serogroups, appearing in just 1-2 studies in either case.

Notice that in terms of countries in the two extremes, we found that Iran in the limit of lowest $\mu$ (co-colonization in fact dominates in carriage) has these unique serotypes not reported in any other studies: 19, 6, 23, 10, 23C, 9, 4B. Whereas, Kenya in the limit of highest ratio of single to co-colonization $\mu$ (single colonization dominates in carriage), has these unique serotypes not reported in any other studies: 47F, 33D, 25A, 5B. 

In each group of studies, any epidemiological setting displays on average about 30 serotypes co-circulating in the same population of children, 32 in A, and 31 in B. When comparing the rank-frequency distributions of serotype occurrences in each group of studies, counting in how many studies (within A and B, respectively) each serotype had been reported, we did not find differences in the rank-frequency distributions for serotype occurrences in each group and when compared to the global dataset (see Supplementary Figs. \ref{fig:rankfreq++}-\ref{fig:plotSerotypeRank}). 

When considering patterns of particular serotype occurrence and co-occurrence, we did not find major differences between groups A and B (Fig.\ref{fig:2networksaccgrAgrB} A-B). In the first group, a total of 92 serotypes are reported, with serotypes shared by all studies being: "14" "19F" "6A" "6B". In the second group, a cumulative number of 80 serotypes are reported, where common serotypes shared by \textit{all} studies without exception are: "6A" "6B" "14" "19F" "23F". In particular, rare serotypes found only in one study comprise about 30\% of serotypes in each group (see Supplementary Table \ref{tab:Serotypes}-\ref{tab:Serogroups} for more details and figures \ref{fig:lessfrequentserotypes}-\ref{fig:uniqueserotypesfrequency}). 

When focusing on those serotypes \textit{common} to both groups of studies, of low and high ratio of single to co-colonization, 70 serotypes in total, including NT (non-typeable), and individually separating those that occur more and less frequently in group A and group B, although we identified some trends for association (Fig.\ref{fig:lowhighmu1}), these did not reach statistical significance. The odds ratios of being reported in either group of studies were not significantly different from 1, for none of the 70 serotypes. However, our procedure could identify for example that serotypes 3, 11, 18,20, 39, 12F, 28F are found in high-$\mu$ settings with probability at least higher than 20\%, compared to low $\mu$ settings. In contrast, in the low-$\mu$ settings, serotypes 37, 10B, 22A, 22F, 33F, 9L, NT are found to occur with probability at least higher than 20\% compared to high $\mu$ settings (see Supplementary Figures \ref{fig:lowhighmu2}-\ref{fig:lowhighmu3}).

When comparing the global probability of co-occurrence in the same site for two serotypes (Fig.\ref{fig:heatmapandSiSjSij}), we consistently find more serotype pairs co-occurring in the same study above chance than below chance, suggesting the existence of some positive mutual feedbacks in serotype coexistence across settings. Furthermore, in this pattern, we do also find a significant difference at the level of $\mu$ groupings (Fig.\ref{fig:2networksaccgrAgrB}C-D): epidemiological sites with low ratio of single to co-colonization (group A: low $\mu$) had relatively less serotype pairs above chance (the product of their occurrences), than sites with higher ratio $I/D$ (group B: high $\mu$). Statistically the odds ratio for serotype pairs co-occurrence above chance vs. below chance, was 2 when comparing group B vs. A ($OR=2.00$, $[95\%CI: 1.79, 2.24]$, Fisher's Exact Test, p-value $<$ 2.2e-16). 

This indicates that sites with high $\mu$ favour more coexistence, and may entail more intrinsic dependencies between serotypes than sites with lower $\mu$, where serotypes appear to co-occur more freely and independently from each other, indicating stronger niche effects. These patterns link with expectations from the mathematical model of \citet{madec2020predicting} with randomly generated interaction coefficients, where $\mu$ is revealed as a gradient for dynamic complexity in coexistence of multiple strains \citep{gjini2020key}. Mathematically, \citet{gjini2020key} show that as a direct consequence of the replicator dynamics at the core, sites with lower $\mu$ (e.g. high $R_0$ for fixed $k$) should be characterized by multi-stability, where fewer serotypes typically coexist, whereas sites with higher $\mu$ (e.g. low $R_0$ for fixed $k$), for the same rescaled interaction coefficients, are characterized by oscillatory inter-dependent dynamics between typically a larger set of serotypes.

\begin{figure*}[!ht]
\centering   
\includegraphics[width=1\linewidth]{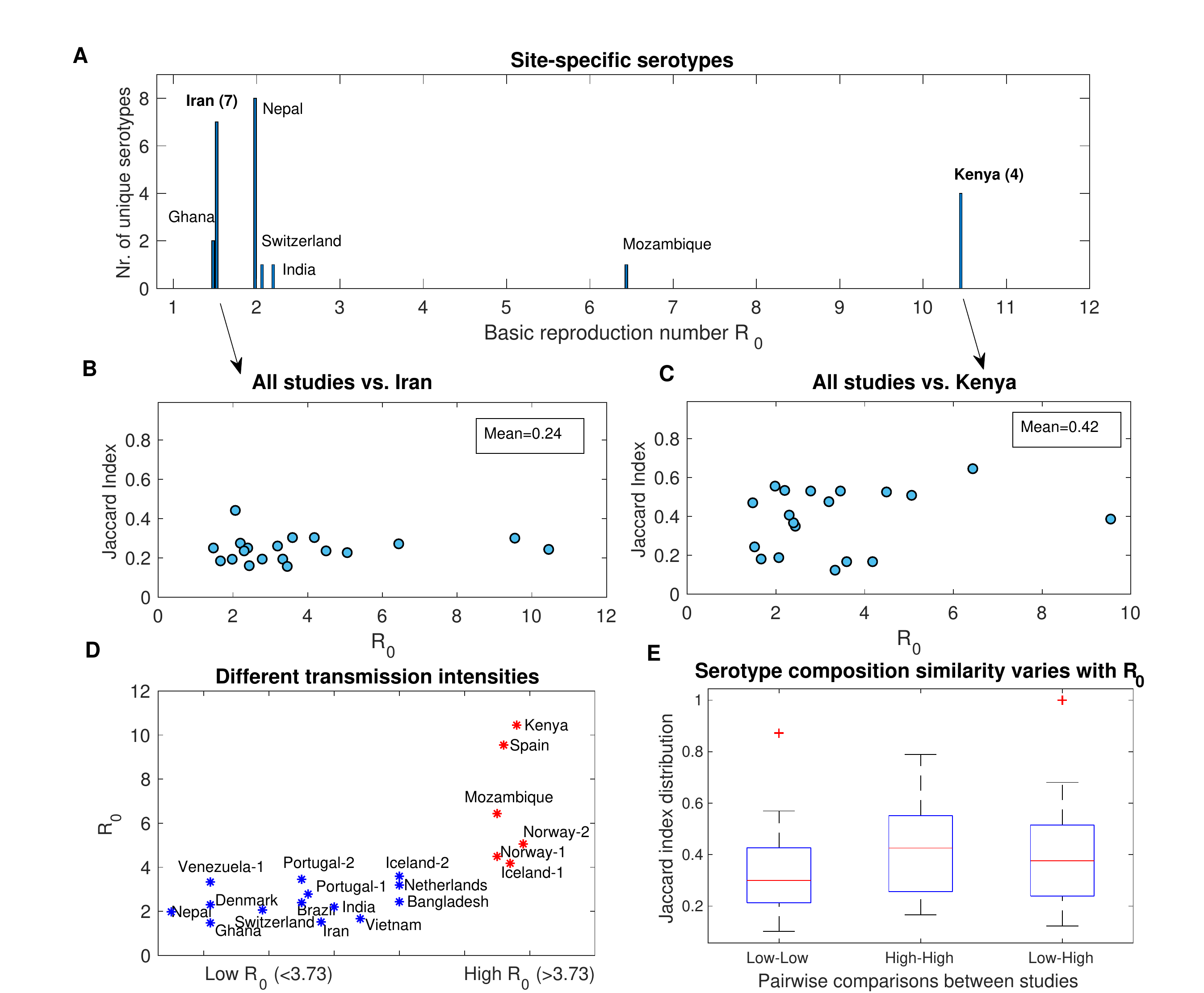}
 \caption{\textbf{Site-specific serotypes, similarity in serotype composition and $R_0$.} \textbf{A. } Here we visualize the number of unique serotypes appearing exclusively in just one study as a function of $R_0$. In Iran (in the limit of lowest $R_0$) the 7 unique serotypes are: "19"  "6"   "23"  "10"  "23C" "9"   "4B", while there are 23 shared serotypes and 72 unique serotypes for the rest of the studies. In Kenya (highest $R_0$) the 4 unique serotypes are: "47F" "33D" "25A" "5B", meanwhile there are 53 shared serotypes and 45 serotypes unique to the rest of the studies altogether.  Iran and Kenya sit in the extremes of the ratio of single to co-colonization $\mu$, lowest for Iran and highest for Kenya. There are few other studies displaying unique serotypes in the dataset, but generally to a smaller degree than these two countries, except Nepal reporting several non-typeable unique serotypes and 8 unique typeable serotypes: "NT4b" "NT2"  "NT4a" "45"   "48"   "NT3b" "11B"  "17A"  "24B"  "32F"  "28A"  "40". India reports serotype 27 as a unique serotype. Mozambique reports "12A" as a unique serotype and Ghana reports "15F" "44" as unique serotypes, not found in other countries. \textbf{B-C.} The values of the Jaccard Index for similarity in serotype composition with Iran and with Kenya, computed for each of the other studies. \textbf{D.} Study distribution according to the values of the basic reproduction number (below and above the mean, in blue and red, respectively 14 and 6 studies). The mean $R_0$ for the first group is 2.46 and the for the second group about 3 times higher, equal to 6.69. \textbf{E.} Distribution of Jaccard indices for pairs of studies according to $R_0$ grouping. The only significant differences are in means of Low-Low (0.32) vs. High-High (0.43) (two-tailed t-test: p-value$<0.05$), and Low-Low vs. Low-High (0.38) (two-tailed t-test: p-value$<0.005$). High $R_0$ implies larger amount of serotype overlap with other studies.}
\label{fig:uniqueSerR0}
\end{figure*}

\paragraph{$R_0$: transmission intensity as a structuring gradient} Finally, we considered the basic reproduction number $R_0$ as a potential key factor in structuring serotype composition, but at a coarser level of resolution than Fig.\ref{Jaccarddheboxplot}D. First, we notice that the epidemiological settings displaying extreme values of transmission intensity, in the limits of low $R_0$ and high $R_0$ have more unique serotypes, that are not shared by other studies in the global dataset. For example, Iran and Ghana with $R_0=1.5$ together harbor 9 unique serotypes: "19"  "6"   "23"  "10"  "23C" "9"   "4B" (Iran), and "15F" "44" (Ghana) meanwhile Kenya with the highest value of $R_0$ ($R_0 >10$) has 4 unique serotypes: "47F" "33D" "25A" "5B". In general, other studies in the dataset share only about 24\% of their serotypes with Iran, but about 40\% of serotypes with Kenya. This pattern of higher serotype sharing with high-transmission scenarios and lower serotype sharing with low-transmission scenarios is persistent and significant even when considered in pairwise comparisons across the entire dataset, if studies are subdivided into \textit{low} and \textit{high} transmission sites, e.g. $R_0$ below and above the mean: $R_0<3.73$ (14 studies) and $R_0\geq 3.73$ (6 studies), as illustrated in Figure \ref{fig:uniqueSerR0}. 

Hence there are two phenomena to note in this final finding and relate with theory: i) sites with low $R_0$ are intrinsically prone to more stochasticity, because of lower overall prevalence of carriage, and hence larger demographic fluctuations should create variability in surviving serotypes, ii) sites with low $\mu$ are intrinsically increasingly characterized by multi-stability and hence could manifest sets of non-overlapping coexisting strains \citep{gjini2020key}. When low $R_0$ settings also have a low ratio of single to co-colonization $\mu$ (e.g. Iran), these two processes may act in synergy, amplifying the uniqueness and variability of coexisting serotypes. Similarly, high $R_0$ - high $\mu$ settings (e.g. Kenya) should show the opposite pattern: on one hand, high prevalence reduces demographic stochasticity, hence extinction probabilities, and thus allows more serotypes to survive, and high $\mu$ promotes also a larger number of serotypes coexisting because of the nature of their intrinsic coupling via mutual interactions in co-colonization \citep{gjini2020key}. 

Measuring pneumococcal co-colonization in natural host populations is challenging, and various methods exist with more research ongoing to optimize detection and quantification of multiple serotypes and strains \citep{shak2013influence,leung2012sequetyping}. Existing estimates from different studies suggest large variation across geographic and demographic settings, but it is important to keep in mind that any isolated reports of co-colonization prevalence are not informative by themselves without the complete backdrop of single and total colonization prevalence. Beyond statistical approaches, a general mechanistic and simple enough epidemiological model is needed to interpret such numbers and make sense of their joint variation, providing them with biological and theoretical nuance. While full mechanistic characterization of all diversity found in pneumococcus may be an out-of-reach task, we propose that many features of the system and dynamics can be modelled and understood using the quasi-neutral SIS model with co-colonization and multiple interacting strains \citep{madec2020predicting}, and eventually, its more recent extensions and generalizations \citep{le2021quasi,le2022disentangling,le2021reaction}. One attempt to apply such model to pneumococcus global epidemiological data has been made here, giving validation to the hypothesis that environmental adaptation drives pneumococcus co-colonization prevalence. Using this explicit modeling framework, many more qualitative and quantitative spectra of the dynamics of pneumococcus serotypes, including in response to antibiotics or vaccine pressure, could be modelled and investigated in the future.

\section{Discussion}
Our work is intended first, as an initiative to develop an integrated understanding of pneumococcal co-colonization variation and its significance around the world, currently regarded as elusive, and secondly as a fresh exploration of the stress-gradient-hypothesis in a new mathematical and ecological context.
There has been substantial research on pneumococcus competitive abilities in multiple colonization in recent years, and on quantifying the major, although not exclusive, role of capsular serotype for their determination as a bacterial phenotype \citep{lipsitch2012estimating,trzcinski2015effect,weiser2018streptococcus,abruzzo2022serotype}. Yet, an ecological understanding and integration of all the findings is far from complete. 
To our knowledge, epidemiological surveys on the prevalence of co-colonization by multiple serotypes of pneumococcus have not been comparatively assessed and modelled, the focus having been mainly on describing trends in global invasive disease \citep{o2009burden,song2013clinical} or patterns of antibiotic resistance variation \citep{reinert2005antimicrobial,davies2019within}. A few epidemiological modeling studies perform comparative analyses between a few countries to distinguish epidemiological parameters responsible for carriage differences or serotype replacement post-vaccination (e.g. \citep{pessoa2013comparative,gjini2017geographic}, although the need for making sense of differences in pneumococcal epidemiology across geographical settings is widely recognized \citep{chan2019using,lewnard2019making}.

In this paper we address this challenge, comparing colonization and co-colonization rates in multiple epidemiological settings from different geographical locations, within the framework of a simple, yet general enough mathematical model \citep{madec2020predicting}, and linking these data to the ecological question of the stress-gradient-hypothesis \citep{bertness1994positive}: Do pneumococcus competitive abilities, when viewed as a whole, vary with transmission intensity? 

The SGH predicts that higher intensity of stress, in our case, lower overall carriage prevalence,  should favour increased facilitation between ecological members of the communty, in our case higher susceptibility to co-colonization among serotypes, and viceversa, higher carriage prevalence should promote more competition and hence a lower susceptibility to co-colonization among serotypes. The $N$-strain epidemiological model proposed by \citep{madec2020predicting} makes this link entirely explicit, because it allows us to quantify these two parameters independently of each other, and interpret the SGH as a critical trade-off that preserves (mathematically and biologically) a given stability-diversity-complexity configuration in multi-strain coexistence \citep{gjini2020key}.

By quantifying co-variation patterns between $R_0$ and $k$ across epidemiological settings, we found a signature of the stress gradient hypothesis in the global epidemiology of pneumococcus. Although there is variation in $R_0$ and $k$ across locations, these two epidemiological parameters inversely co-vary, giving rise to a conserved single to co-colonization ratio $\mu=I/D=[k(R_0-1)]^{-1}$ as predicted by the model \citep{gjini2020key}, which for our data emerged to be 6.03 (95\%CI: 3.56,10.21) close to the geometric mean of empirical single-to-cocolonization ratios reported globally. Furthermore, we demonstrated that even when stratifying the data further by $\mu$, at higher resolution (e.g. two clusters of epidemiological settings, low and high ratio of single to co-colonization ( $\mu\approx 3$ vs. $\mu\approx 12$), or even three clusters as in Fig.\ref{fig:3clustSGH}) the same pattern of co-variation between $R_0$ and $k$ is preserved. 

From our mathematical model's simulations (Fig. 5 in \citep{gjini2020key} for $N=10$), in sites within the range of single to co-colonization ratios $\mu$ observed in this dataset (e.g. $\mu \in [3,12]$), the probability of \textit{`no stable steady state'} coexistence exceeds 60\% and is increasing with $\mu$, while the probability of \textit{`a single stable steady state'} is only above 30\% and decreasing, with only a very small chance of multi-stable coexistence among strains, hence suggesting that a likely natural mode of coexistence of pneumococcus serotypes worldwide may be oscillatory dynamics. That is perhaps, why we don't find a clear signature of $\mu$ on the compositional variation across sites. Although compositional variation across local sites in terms of pneumococcus serotype sharing could be related to theoretical expectations on higher beta diversity in more productive environments \citep{chase2010stochastic}, to disentangle the effect of $R_0$ (`productivity') from $\mu$ (under fixed $k$), one would need more representation of sites in the $\mu \to 0$ spectrum, where multi-stability is more likely. In that case, `moving' horizontally from low to high $R_0$ (Fig. \ref{muclust}D), and crossing lines of decreasing $\mu$ ($\mu\leq1$), we should expect to find higher compositional diversity across settings.

We found that our studies falling along the same $\mu$ line, i.e. satisfying the SGH, did not significantly differ in serotype composition, as measured by the Jaccard index, pointing to adaptation of the same set of serotypes by upregulation or downregulation of competitive abilities in different environments. The mechanisms of such adaptability could implicate other loci responsible for intra-host competition. Pneumococcus strains are known to express several quorum-sensing regulated factors that can kill other pneumococci not expressing cognate immunity factors. The \textit{blp} locus encodes a family of bacteriocins (pneumocins), which have been shown to promote competitive interactions between different strains \citep{wholey2019characterization}. In addition, a separate quorum-sensing system causes lineage-independent killing of pneumococci (known as fratricide) that have not achieved the required population size for immune activation also regulated by this system \citep{claverys2007competence}.

Epidemiological studies by nature are observational, and involve strains that differ in many loci apart from the capsule. Thus, although we focus on serotype as the basic unit of diversity in multiple colonization, with the advantage of being able to capture and quantify the relevance of serotype differences in humans, we must keep in mind that a natural feature of the analyzed studies is further diversity underlying each serotype (different genetic backgrounds, strains differing in other loci, for example, antibiotic resistance determinants, blp locus, quorum-sensing, competence, horizontal transfer). Such hidden underlying diversity most likely plays a key role in the adaptability required for the ecological trade-off we uncover. A typical feature of the pneumococcal genome, evidenced by many studies \citep{claverys2000adaptation,straume2015natural}, is its plasticity, which facilitates this bacterium's adaptation to changes in the environment, evolution by lateral gene transfer and recombination \citep{mortier2007key,bruckner2004mosaic}, and has led to a large pangenome \citep{hiller2007comparative,straume2015natural}. 

While stress to the bacteria can come in many forms (antibiotics \citep{mitchell2015carriage}, availability of susceptible hosts, quality of susceptible hosts, transmission constraints, temperature and other climatic conditions \citep{de2019unraveling}), its ultimate manifestation at the epidemiological level is in reducing the basic reproduction number $R_0$, hence endemic prevalence of the pathogen in a given setting. Here, we have shown that despite the heterogeneity of the cross-sectional datasets, there is a consistent pattern of the bacteria increasing their co-colonization efficiency $k$, hence tending towards more mutual facilitation and less competition, in response to stress. This upregulation of co-colonization susceptibility can be a survival-promoting mechanism in harsh environmental conditions, which allows the pneumococci to benefit from the co-presence of other pneumococci, e.g. by taking advantage of the high transformation efficiency during co-colonization by multiple strains \citep{marks2012high}.  Remarkably this trade-off pattern, mirroring also the stress-gradient hypothesis \citep{bertness1994positive,callaway1997competition}, and preserving key qualitative features of multi-strain coexistence \citep{gjini2020key}, holds across different human populations and continents and is not confined to a particular region. 

On average our cocolonization studies display a similar serotype composition, with about 35\% of reported serotypes shared between any two studies, which increases to 50\% when measured at the serogroup level, and does not vary significantly in low vs. high ratio of single to co-colonization settings ($\mu$ clusters). Even though we found some structure in terms of higher or lower mean Jaccard index within the big set of pairwise study comparisons (Figure \ref{fig:distJac2}), starting from Jaccard-index similarity, we did not find significant differences between epidemiological parameters among such clusters (Figure \ref{fig:clustJac2}), supporting that serotype composition is not a primary driving factor in the country-wise differences. However, when starting from transmission intensity similarity, and considering the studies in terms of low $R_0$ and high $R_0$, we found that sites with low $R_0$ share less serotypes than sites with high $R_0$, suggesting that high transmission intensities, and hence levels of overall carriage prevalence, likely support more similar serotype coexistence. 

Despite geography being a natural factor in the data set, since our studies reflect different populations, and environmental and climatic conditions, when searching for associations between geographic distances and epidemiology, the geographic distance analyses do not support a significant association with differences in epidemiology, e.g. at the level of countries in Europe vs. rest of the world ($\mu$ comparisons in Fig.\ref{fig:geoheatmap}), and in general no significant association between ($R_0, k$) distances and geographical distance. However, we did find a significant, albeit weak, trend of lower serotype composition similarity (Jaccard index) with increasing physical distance between any two study locations (Fig.\ref{fig:geoscatJ}). This is similar to what has been found before in a study of sub-saharan Africa where close geographic proximity ($<5$km) between shared pneumococcus strains was associated with a significantly lower pairwise SNP distance compared to strains shared over longer distances \citep{senghore2022widespread}. Associations with geographic proximity would need to be explored model extensions that account explicitly for host movement and serotype transmission in space, and remains an interesting avenue for the future.

Our modelled dataset excluded certain studies which belonged to a longitudinal design (Finland \citep{syrjanen2001nasopharyngeal}, Peru \citep{nelson2018dynamics}, South Africa \citep{manenzhe2020characterization}, Gambia \citep{chaguza2021carriage}), and some cross-sectional studies that were deemed as outliers in our preliminary tests in Greece \citep{syrogiannopoulos2002antimicrobial} and Venezuela-2 \citep{rivera2011carriage}. More recently, we became aware of some other very early studies, such as a study of co-colonization in Australian children \citep{hansman1985pneumococcal} and in Papua New Guinea \citep{gratten1989multiple,montgomery1990bacterial}, and another study that only recently came out on pneumococcus carriage patterns in Poland \citep{wrobel2022pneumococcal}. Reassuringly, when superimposing such studies, on our model-fitted data set, in terms of their $R_0$ and $k$ estimates (see Table \ref{tab:Otherstudies} and Figure \ref{fig:otherpred}), we observe that the new cross-sectional studies do seem to fall within the same linear trend predicted by our model, consistent with the SGH, while the longitudinal studies seem to deviate more. Such deviation could be related to a higher frequency of repeated measurements in the same children in longitudinal studies, which could bias estimates. 

An exciting avenue for the future would be to continue to gather new cross-sectional studies and examine their reports for single and co-colonization prevalence of pneumococcus serotypes under the same integrated modeling lens. Exploring differential adjustment in co-colonization efficiency in response to transmission intensity variation over time and space, possibly over smaller scales, (e.g. same country over different seasons) could also bring new information to the picture of pneumoccocus adaptation, similar to what has been observed for malaria parasites \citep{oduma2021increased}, where it was found that parasites increase their investment in transmission in the wet season, reflected by higher gametocyte densities, even though in the wet season, fewer infections harbored detectable gametocytes. 

Other promising directions would be to dig deeper into patterns of serotype frequencies, dominance and variation in particular epidemiological and geographical settings, and link those to the features of equilibrium and non-equilibrium coexistence expected from the quasi-neutral SIS model with replicator dynamics at its core \citep{madec2020predicting, gjini2020key}. While not all studies analyzed here report serotype frequencies, a few of them do, including Denmark \citep{harboe2012pneumococcal}, Iran \citep{tabatabaei2014multiplex}, Brazil \citep{rodrigues2017pneumococcal} and Mozambique \citep{adebanjo2018pneumococcal}, encompassing the spectrum from lower to higher ratios of single to co-colonization.
Alternatively, one could try to revisit such data in light of more general macroecological laws that have been proposed to describe variation and diversity in microbial communities, based on the stochastic logistic model and environmental fluctuations \citep{grilli2020macroecological}. Developing more standardized model-data links for inference and prediction, from a global standpoint or on a case-by-case basis, remains a challenge for a deeper understanding of pneumococcus diversity, coexistence dynamics and control. 

\section{Acknowledgement}
This study was supported by the Portuguese Foundation for Science and Technology (FCT) in Lisbon, Portugal (grant number LISBOA-01-0145-FEDER-029161) and Instituto Superior Tecnico, via fellowship BL183/2021/IST-ID to ED. The Center for Stochastic and Computational Mathematics is supported by FCT via UIDB/04621/2020 and UIDP/04621/2020.

\section{Supplementary material (online)}
S1: Data and notes on data collection, analysis and preparation. \\
\noindent S2: The full SIS mathematical model for co-colonization. \\
\noindent S3: Supplementary figures and tables.

\newpage
\begin{center}
\large{\underline{\textbf{Supplementary materia}l}\\
Pneumococcus and the stress-gradient hypothesis: a trade-off links $R_0$ and susceptibility to co-colonization across countries}\\
\normalsize
{Ermanda Dekaj $^{1,\dag}$, Erida Gjini $^{1,*}$ \\
\small{$1$ Center for Computational and Stochastic Mathematics, Instituto Superior Tecnico, University of Lisbon, Lisbon, Portugal}\\
\small{ $^\dag $ermanda.dekaj@tecnico.ulisboa.pt,$^*$ Corresponding: erida.gjini@tecnico.ulisboa.pt}}\\
\end{center}

\setcounter{figure}{0}
\setcounter{section}{0}
\setcounter{table}{0}
\setcounter{page}{1}
\renewcommand{\thefigure}{S\arabic{section}}
\renewcommand{\thefigure}{S\arabic{figure}}
\renewcommand{\thesection}{S\arabic{section}}
\renewcommand{\thetable}{S\arabic{table}}

\section{\underline{Notes on data analysis and code availability}}

\paragraph{Data Collection Characteristics of table 1} 
The summary of the reported values for epidemiological quantities of pneumococcal colonization and co-colonization can be found in the Table 1. More extensive details are found in the Supplementary Data file: \underline{$MetadataSGHSpn.xlsx$}. \\

\par The Iran study \cite{tabatabaei2014multiplex} reports rates for T, D and sample size $n$, from where we calculate I. They report the total number of serotypes they detected (30). We get the information from table 3 of the article \cite{tabatabaei2014multiplex}.

\par In the Spain study \cite{ercibengoa2012dynamics}, 25 children were sampled three times, 53 children twice and 27 children once, over 12 months. We use the rate they report for colonized children from the total of 105. They report 21 serotypes, as many as they detected. We get the information from table 2 of the article \cite{ercibengoa2012dynamics}.

\par In the Venezuela-1 study \cite{rivera2009multiplex}, they reported 50 primary cultures from the nasopharyngeal swabs of healthy Warao-Amerindian
children. They report serotypes detected on multiple carriers in Table 1 of the article \cite{rivera2009multiplex}. These are the serotypes we include in our serotype composition data for this setting. 
Venezuela-1 and Venezuela-2 report the rates from children in Warao and Caracas city respectively.

\par In the Iceland study \cite{hjalmarsdottir2016cocolonization} they compare two methods for detecting multiple carriage: 1) conventional and 2) molecular, whose results we keep in our meta-analysis as Iceland-1 and Iceland-2 respectively. We split this study in two, because they do not report a final single value for overall co-colonization percentage. In contrast, other studies, e.g. Netherlands \cite{wyllie2016molecular} and Brazil \cite{rodrigues2017pneumococcal}, also compare two and three methods respectively, but they do report the overall percentage, which we then include as a single value in our metadata.

\par In the Netherlands study \cite{wyllie2016molecular}, the number of carriers of multiple serotypes of Streptococcus pneumoniae os reported via both methods molecular and conventional. We get the overall rate as reported in the study by consideration of both methods. They report serotypes as 6A/6B, and we consider them as two serotypes 6A and 6B for our analysis. They report pneumococcal serotypes detected only by the conventional method in all 803 carriers of pneumococci and we retrieve the overall number of serotype strains detected by conventional culture and serotype-specific signals detected by molecular method (qPCR), among all infants identified as carriers of S. pneumoniae by either method used in the study. 

\par In the Bangladesh study \cite{saha2015detection}, they report multiple serotypes carried from some of the nasal swabs of the children carrying pneumococcus. We calculate the expected proportions of single and double co-colonization while maintaining the rates of I and D reported for those samples analyzed carrying pneumococcus. In the Bangladesh study, the samples were collected in a government housing area. 

\par In the Nepal study \cite{kandasamy2015multi}, they report the colonization and co-colonization rates for typable pneumococci. 

\par In the Portugal study \cite{valente2012decrease} \cite{sa2009changes} the prevalence of pneumococcus carriage is reported for 2001 (before the vaccine introduction) and in 2006 (after vaccine introduction where most of hosts are vaccinated). Since Valente et al. \cite{valente2012decrease} chose only a subset of the samples carrying pneumococci for analysis of multiple carriage, we extrapolate the calculations for all children in their overall sample, by preserving the same percentages of single and double colonization. We have coded Portugal-1 as the epidemiological setting before PCV7 vaccine introduction, and Portugal-2 as the data in the post-vaccine era, where most of the children are vaccinated. These data were also used in previous modeling studies \citep{gjini2017geographic}. Serotypes of Portugal-1 and Portugal-2 were retrieved from \cite{valente2012decrease}.

\par In the Norway study \cite{vestrheim2010impact}, the authors report the rate for single and double colonization. Norway-1 corresponds to non-vaccinated children in 2006 and Norway-2 to vaccinated children in 2008. Serotypes were retrieved from the table reported in article for both periods respectively. These data have also been used in a previous modeling study \citep{gjini2017geographic}. 

\par Similarly the India study \cite{sutcliffe2019nasopharyngeal} reports single and double colonization with S. pneumoniae serotypes, before vaccine introduction, with the samples collected in December 2016 to July 2017, from healthy community children. Children in the community were enrolled from one of the primary care clinics, the immunization clinic at the district hospital, and several
(daycare) centers. Serotypes are taken from figure 1 of the article.

\par In the Switzerland study \cite{brugger2009detection}, they report colonization rates and co-colonization rates according to some of positive samples carrying pneumococci. Data belong to a wider nationwide surveillance program. We perform the calculations for expectations regarding the epidemiological status of all children in the big sample size, while maintaining the same colonization and co-colonization rates detected in the subset of samples analyzed, by the PCR method. Children attending and not attending day-care were part of the sample, with co-colonization not being associated with more often attending day care. 

\par In studies from Brazil \cite{rodrigues2017pneumococcal}, Denmark \cite{harboe2012pneumococcal}, Vietnam \cite{dhoubhadel2014bacterial}, we obtained the rates they reported for colonization and co-colonization, and serotypes from the text. 57\% of samples in Brazil study were collected in one hospital and the others from one childcare centre. 
 
\par In the Mozambique study \cite{adebanjo2018pneumococcal} they reported the rates of colonization and co-colonization, for children with and without pneumonia. We included in our metadata just the rates reported for children without pneumonia. They report the dataset for all serotypes detected.

\par In the Ghana study \cite{dayie2013penicillin} they reported the overall rates for colonization and co-colonization collecting data from two cities Accra and Tamale. In the Ghana study, children were attending nurseries and kindergartens.

\par In the Kenya study \cite{kobayashi2017pneumococcal} the serotypes were retrieved from the text and from the additional supplementary material accompanying the paper.

\paragraph{Data file 2} Serotypes reported in each study and statistics of occurrence/co-occurrence, and rankings in the global dataset. This file presented with a matrix is used to analyse the occurrence/co-occurrence of serotypes among different countries. Each row corresponds to a specific serotype and their co-occurrences with other serotypes. In the diagonal we report the number of studies each serotype was reported. We use these data for ChordDiagram visualizations and all analyses regarding serotype composition similarity across studies. File: \underline{$SerotypesSGHSpn.xlsx$}

\newpage

\section{\underline{The full SIS mathematical model for co-colonization dynamics and $N$ strains}}
We apply a model for epidemiological transmission of $N$ strains, following \textit{Susceptible-Infected-Susceptible} dynamics, with coinfection, described in detail in \citep{madec2020predicting}. The variables are: $S$, $I_i$ and $I_{ij}$, referring respectively to the proportion of uncolonized hosts, those singly-colonized by strain $i$ and those co-colonized by strains $i$ and $j$. How these variables change over time is given by the following system of ordinary differential equations:
\begin{eqnarray}\label{system0}
\frac{dS}{dt}&=&m(1-S)-S\sum_{j=1}^N F_j,\\
\frac{dI_i}{dt}&=&F_i S-mI_i- I_i \sum_{j} K_{ij}F_j ,\quad 1\leq i\leq N\\
\frac{I_{ij}}{dt}&=&I_i  K_{ij}F_j -m I_{ij}, \qquad 1\leq i,j\leq N
\end{eqnarray}
where $F_{i}=\beta \left(I_i+\sum_{j=1}^N \frac{1}{2}(I_{ij}+I_{ji})\right)$ gives the force of infection of strain $i$ in the system, summing the contribution of all hosts transmitting that strain, at transmission rate $\beta$. Using $S=1-\sum_{i=1}^N{I_i} -\sum_{i,j}I_{ij}$, the dimension of the system is effectively $N+N(N-1)/2$. Co-colonized hosts with strains $i$ and $j$, transmit either with equal probability 1/2. In the system, the parameter $m=\gamma+r$, depicts infected host turnover rate, summing the effects of both the clearance rate $\gamma$ of colonization episodes and the recruitment rate of susceptibles $r$. Susceptible birth rate is assumed equal to natural mortality rate. 

Upon co-colonization, strains can interact with each-other, and such pairwise coefficients are represented by the matrix $K$, where values $K_{ij}$ above 1 indicate pairwise facilitation, while values of $K_{ij}$ below 1 reflect competition in co-colonization between $i$ and $j$, namely a prior colonization by strain $i$ reduces the susceptibility of the host to co-colonization by strain $j$. In \citep{madec2020predicting}, a slow-fast timescale decomposition of the dynamics has been derived, assuming strain similarity in co-colonization interaction coefficients. 
Mathematically we have: $$K_{ij}=k+\epsilon \alpha_{ij},$$ where $\epsilon$ is small ($0<\epsilon \ll 1$), and $k$ is a convenient reference, e.g. the mean of  $K_{ij}$: $k=\frac{\sum_{1\leq i,j \leq N} K_{ij}}{N^2}$. This yields a normalized interaction matrix $$A=(\alpha_{ij})=\frac{K_{ij}-k}{\epsilon}$$ with the same distribution as $K$, but with mean 0 and variance 1 (if $\epsilon$ is taken to be the standard deviation of $K$), at the start of dynamics. 

With this formulation, the $N$-strain frequency dynamics are neutral over the fast timescale $o(1/\epsilon)$ but over the slow timescale $\tau=\epsilon t$ are captured by the replicator system:
$$\frac{d}{d\tau} z_i = {\Theta z_i \cdot\bigg( \sum_{j\neq i} \lambda_i^j z_j -\mathop{\sum}_{1\leq k<j\leq N} (\lambda_j^k+\lambda_k^j) z_jz_k \bigg)},\quad i=1,\cdots,N$$
where $\lambda_i^j=\alpha_{ji}-\alpha_{jj} -\mu(\alpha_{ij}-\alpha_{ji})$ denote pairwise invasion fitnesses between any two strains, and $\Theta=\beta\left(1-\frac{1}{R_0}\right)\left( \frac{\mu}{2(\mu+1)^2-\mu}\right)$ gives the speed of the dynamics. The fast variables instead, are the total prevalence of susceptible hosts $S(t)$, singly-colonized hosts $I(t)$, and co-colonized hosts $D(t)$, which reach their equilibrium $(S^*, I^*, D^*)$ on the fast timescale and stay constant over the slow-timescale (see \citep{madec2020predicting,gjini2020key}). It is the analytic expression for this endemic equilibrium that forms the basis of the parameter estimation ($R_0, k$) in this paper.

\newpage
\section{\underline{Supplementary figures}}

\begin{figure*}[!h]
\centering 
\includegraphics[width=0.99\linewidth]{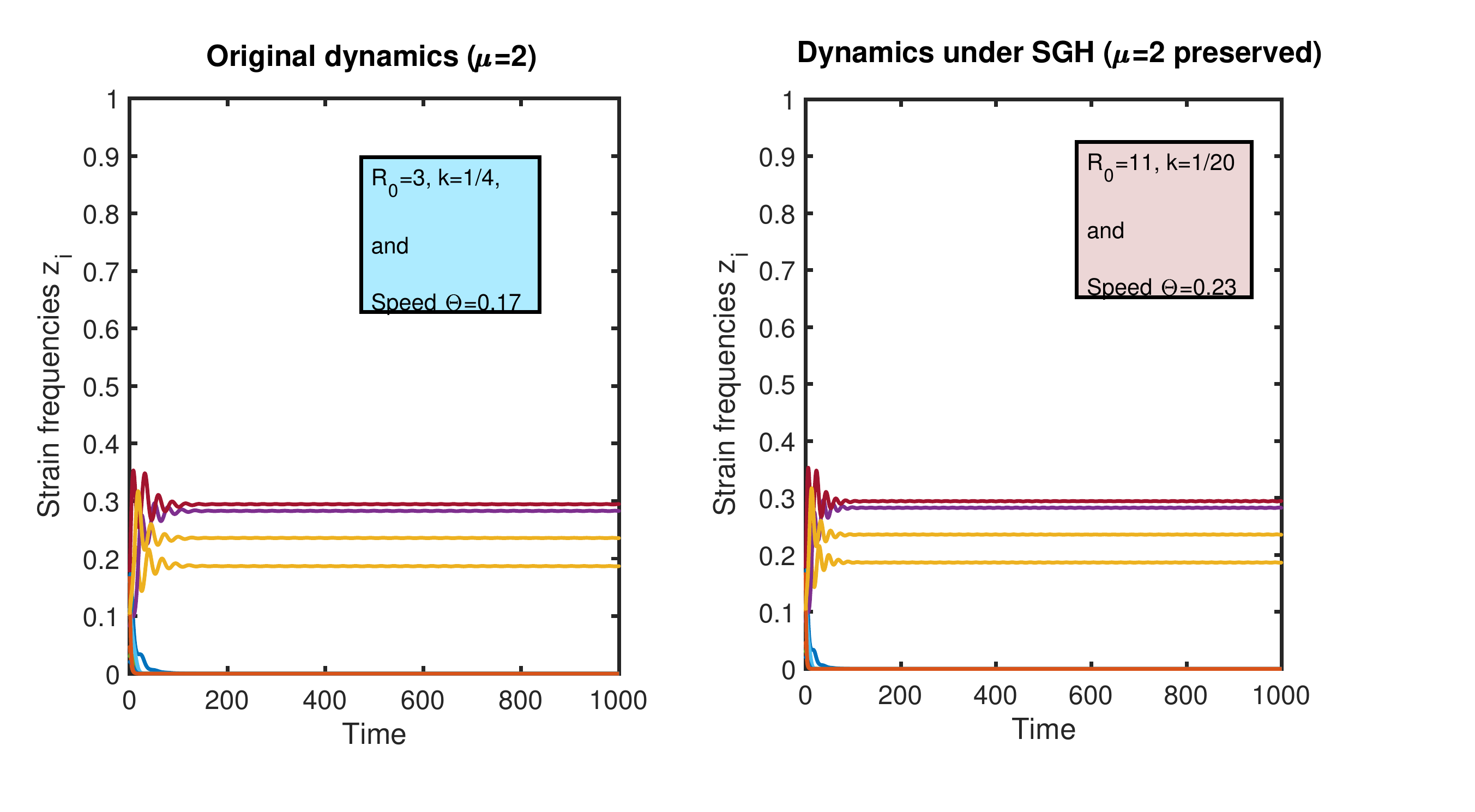}
\caption{\textbf{Preserving the same $\mu$ under global environmental shifts.} We plot dynamics predicted by \citep{madec2020predicting} under the same A matrix as in Figure \ref{fig:model} and varying the two parameters $R_0$ and $k$ in a trade-off manner, so that the mutual invasion fitnesses between system members remain unchanged. Although the qualitative coexistence dynamics is preserved for equal $\mu$, quantitatively the speed of the selection may change with global parameters, individual parameters $R_0$ and $k$ (in this case increasing slightly for higher $R_0$.}
\label{fig:dynSGH}
\end{figure*}

\begin{figure*}[!h]
\centering 
\includegraphics[width=0.99\linewidth]{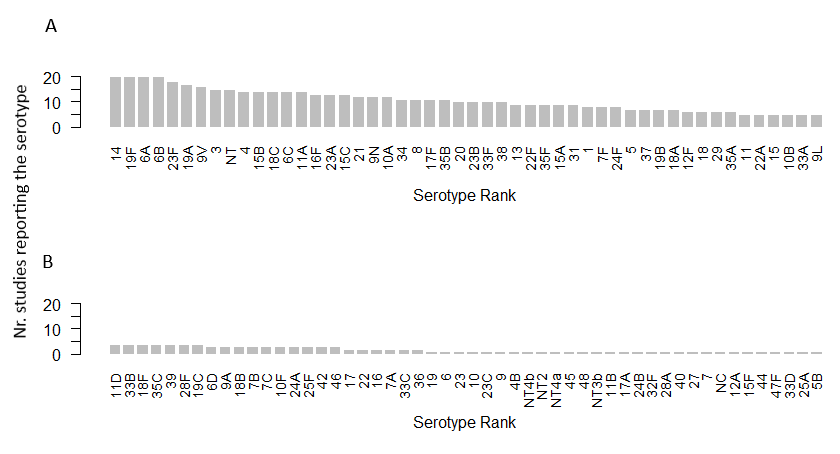}
\caption{\textbf{Serotype occurrences in our dataset of 20 epidemiological studies. } Serotypes are ranked according to the number of studies in our dataset where they were reported, varying from a maximum of 20 studies to a minimum of 1, i.e. from common to rare serotypes globally.}
\label{fig:serotypes}
\end{figure*}

\begin{figure*}[!h]
\centering
\includegraphics[width=0.6\linewidth]{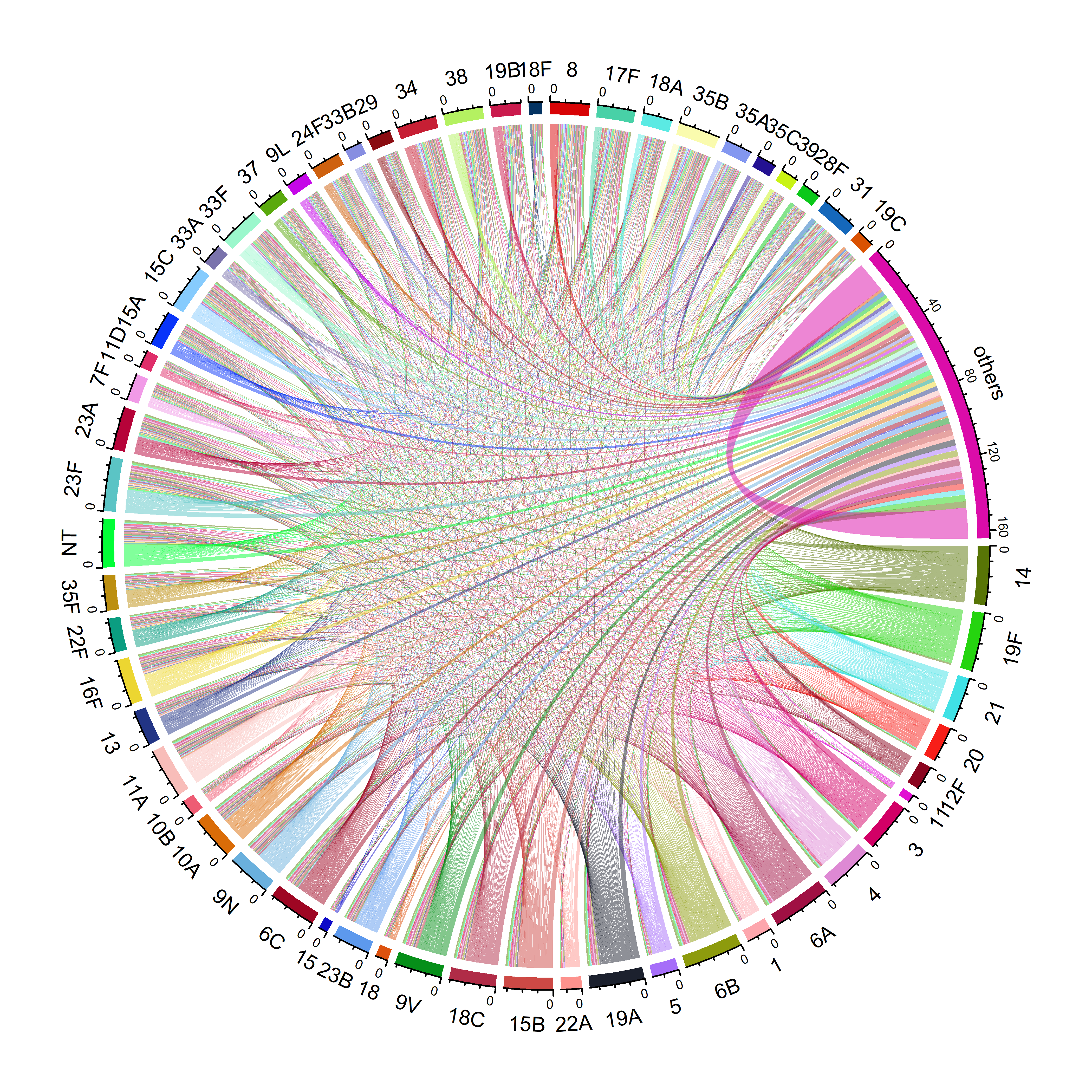}
\caption{\textbf{Serotype composition and co-occurrence across all epidemiological studies.} Network visualization for the pneumococcus serotype set composition across all 20 epidemiological studies from different geographical locations. Here we considered as "others" rare serotypes, those that appeared in only 1, 2 or 3 studies. There is a node for every serotype reported in any of the 20 studies. The thickness of the edge indicates how often serotype $i$ and $j$ were jointly reported in the same study, i.e. found in the same location/population of children. 44.1\% of serotypes (45/102) appeared in only 1, 2 or 3 studies.}
\label{fig:bignetwork}
\end{figure*}

\begin{figure*}[!h]
\centering 
\includegraphics[width=0.7\linewidth]{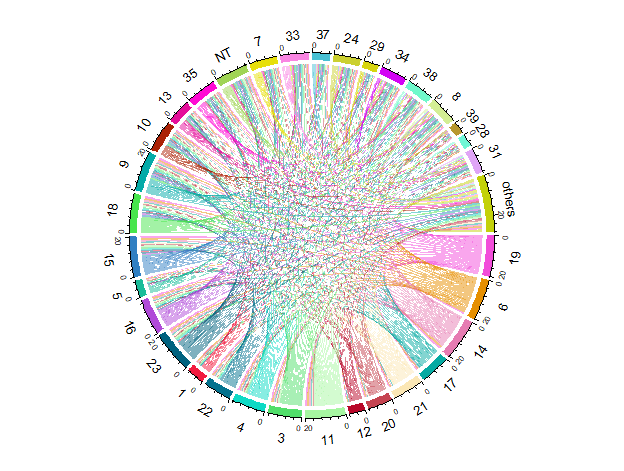}
\caption{\textbf{Serogroup composition across all studies}. Here we see all serogroups reported in all studies, where as "others" are noted serogroups reported in 1, 2 or 3 studies. It results that these subgroup ("others") belongs to 25\% of reported serogroups.}
\label{fig:serogr1}
\end{figure*}

\begin{figure*}[!h]
\centering 
\includegraphics[width=0.9\linewidth]{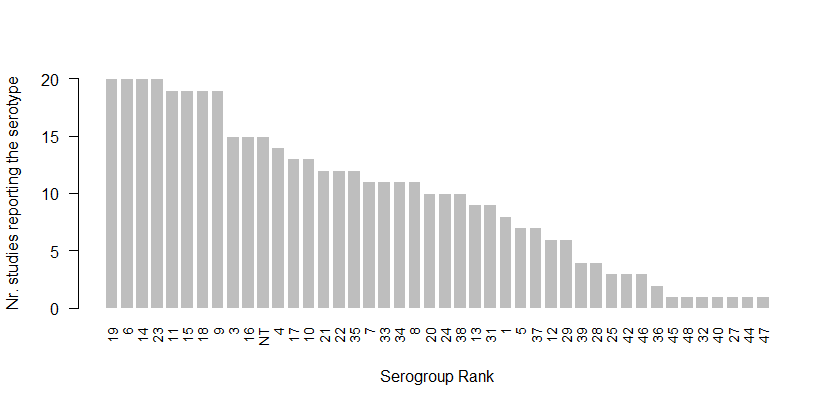}
\caption{\textbf{Serogroup rankings among all studies, according to number of times of occurrence. } Data as in Table \ref{tab:Serogroups}.}
\label{fig:serogr2}
\end{figure*}

\begin{figure*}[!h]
\centering 
\includegraphics[width=1.1\linewidth]{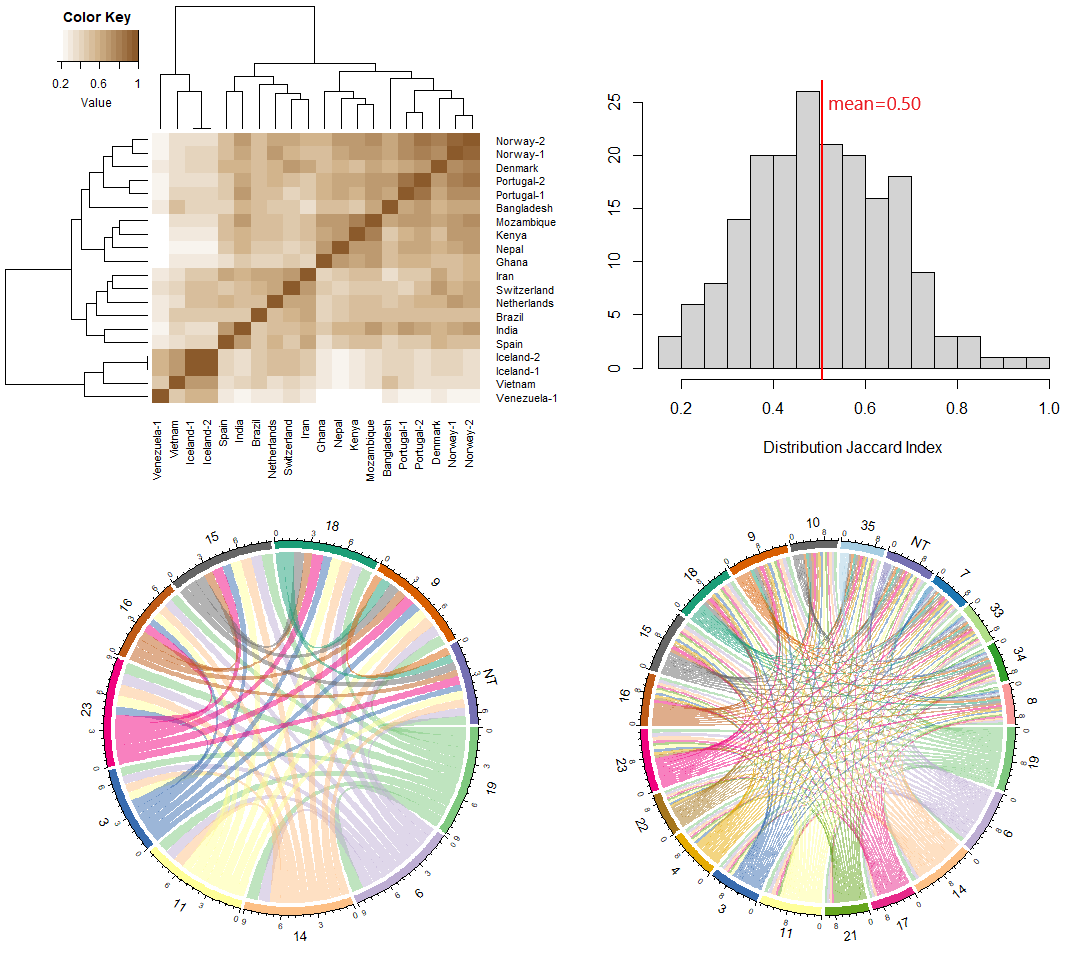}
\caption{\textbf{Serogroup composition similarity across studies} Here we see the serogrups reported in 75\% and 25\% of the studies. On average two studies share 50\% of the serogroups.}
\label{fig:serogr3}
\end{figure*}

\begin{figure*}[!h]
\centering 
\includegraphics[width=0.7\linewidth]{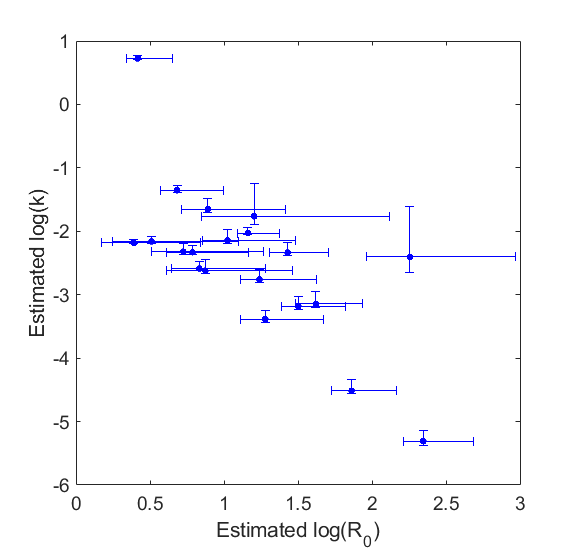}
\caption{Here we visualize the error bars for $R_0$ and $k$ from the confidence intervals calculated using a multinomial distribution with sample size, $N_S$, and the probabilities of $S$, $I$, $D$.}
\label{fig:doubleerrorbar}
\end{figure*}

\begin{figure*}[!h]
\centering 
\includegraphics[width=1\linewidth]{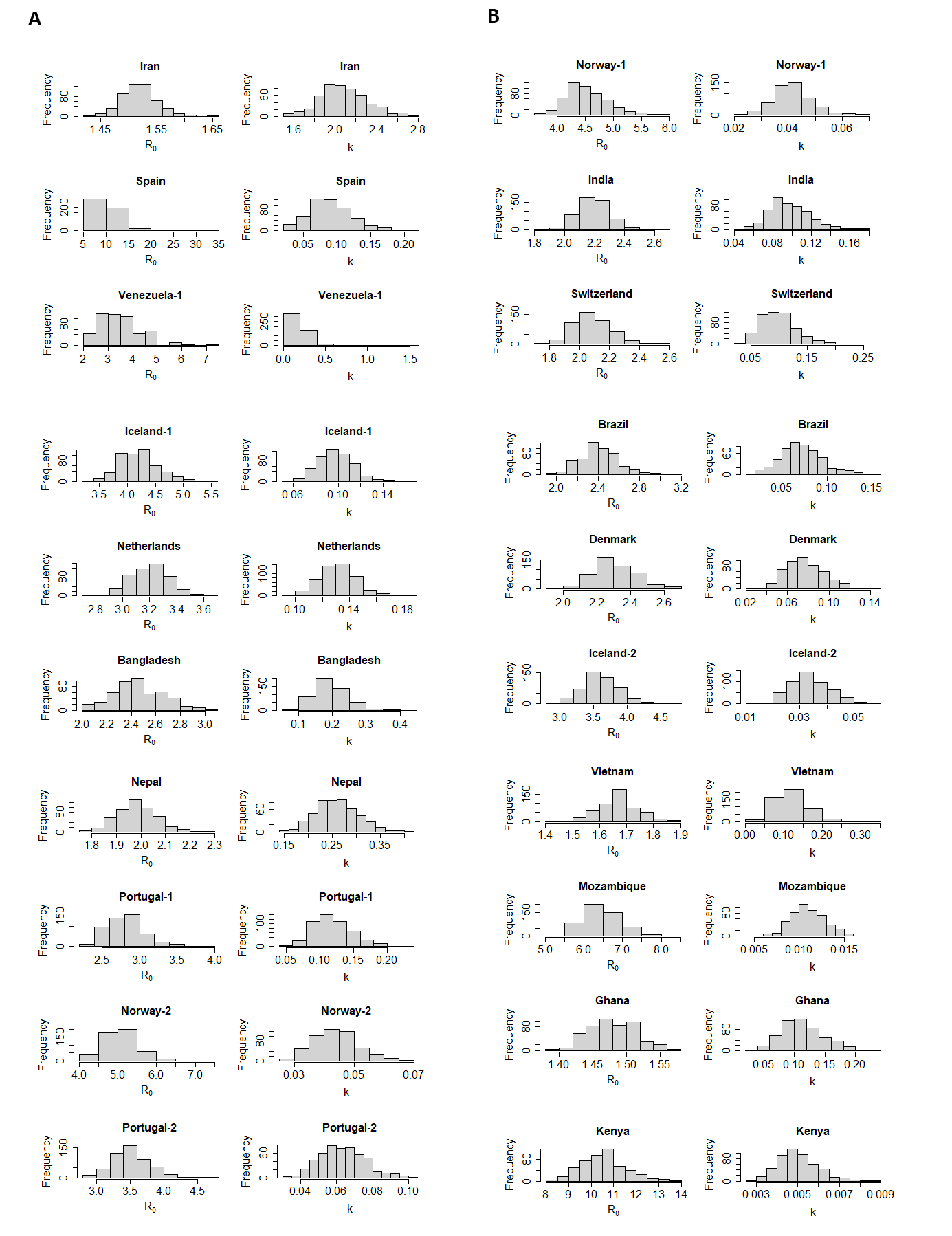} 
\caption{Distribution of all $R_0$ and k simulated (500 simulations) for all countries in order to get the confidence intervals reported in the table \ref{estimates}.}
\label{fig:hists}
\end{figure*}

\begin{figure*}[!ht]
\centering
\includegraphics[width=\linewidth]{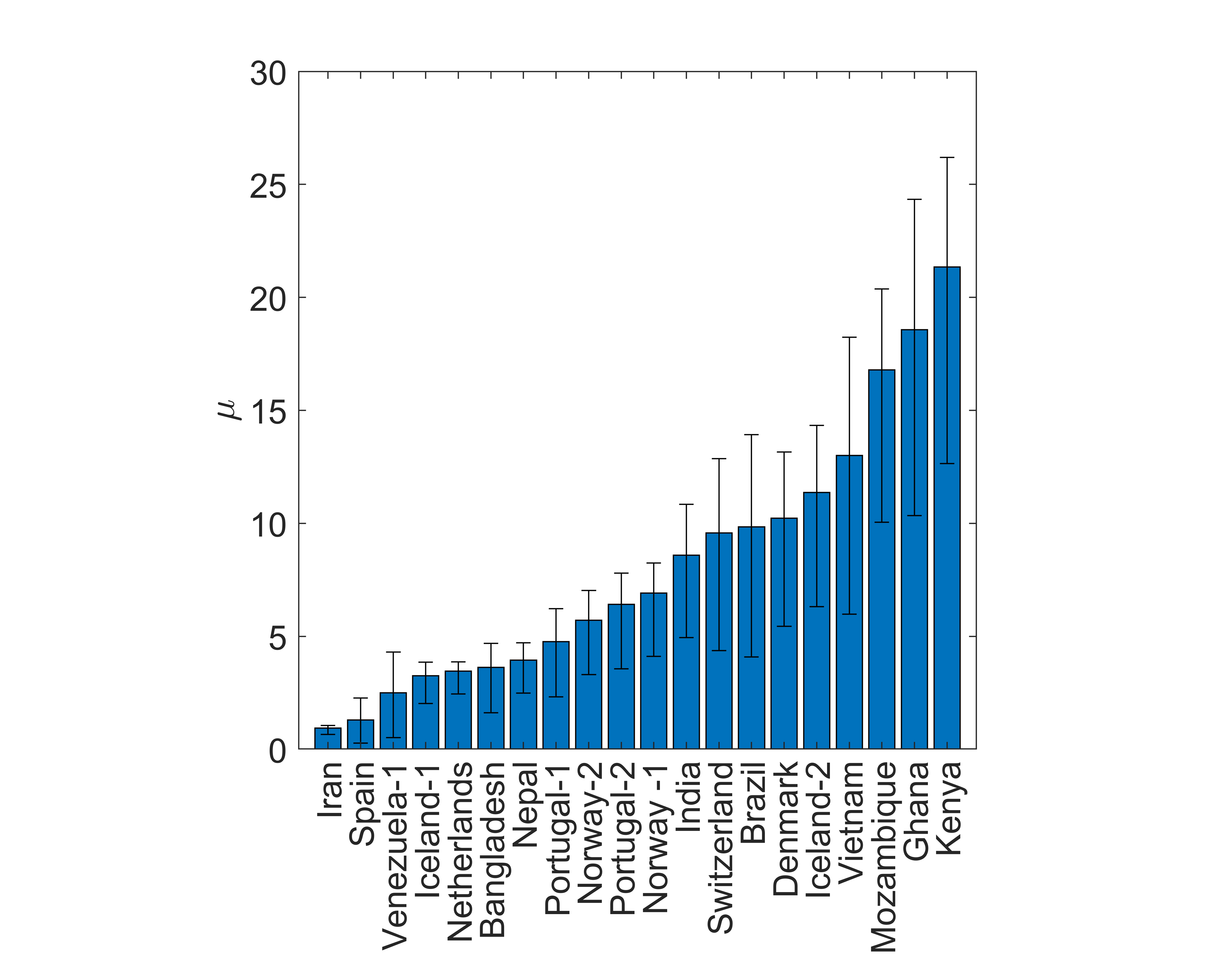} 
\caption{\textbf{Empirical single-to-colonization ratios and uncertainty due to sample sizes.} Here we see the error bar (95\% CI) for each value of $\mu$ for all data, obtained from multinomial resampling (500 realizations) of the datasets using the same frequencies of susceptible ($S$, singly-colonized ($I$) and co-colonized individuals ($D$) reported in each country.)}
\label{fig:empiricalmu_err}
\end{figure*}

\begin{figure*}[!ht]
\centering
\includegraphics[width=\linewidth]{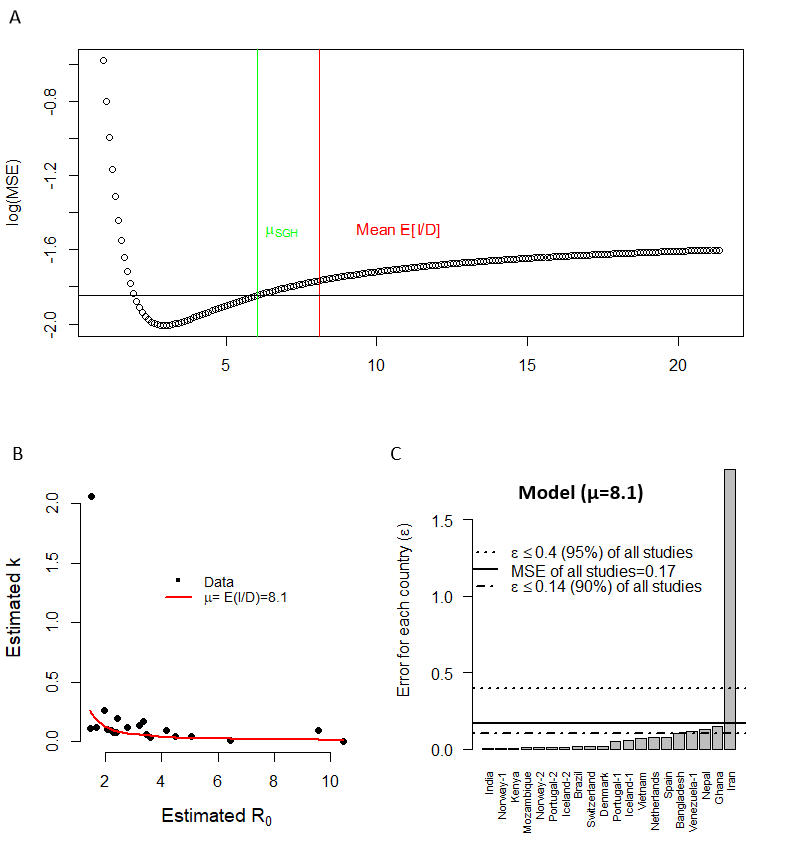}
\caption{\textbf{Examining SGH via a crude optimization approach.} \textbf{Top panel} We can vary $\mu$ systematically over a hypothetical range and compute the SGH-model-predicted $k$ according to the function $f(x,\mu)= \frac{1}{\mu (x-1)}$, where $x = \hat{R_0}$ and $\mu$ is the model's parameter. We did generate $\mu$ values starting from minimum to the maximum of $\mu$ from all the data, by steps of 0.1 ($\mu \in [0.93...21.3]$), then replaced these values in the model-expected $k$, and calculated for each $\mu$ the error between model-predictions and the actual estimates of $k$ for each country, as:
$\epsilon_j(\mu)$=$|k_j-f(x_j,\mu)|$, where j $\in$ data points.
We calculate the mean squared error, for each supposed value of $\mu,$ as: $MSE(\mu) = \frac{\sum_{j=1}^{20} (\epsilon_j(\mu))^2}{20}$. We visualize the error bar for each value of $\mu$ for all data. Examining this quantity as a function of $\mu$ we find that the smallest values of MSE corresponds to $\mu = 2.93$, where we have the minimum value of MSE=0.13. This is very close to the value of mode of $\mu$ (the mode is 3.54). Within 17\% from the minimal MSE we find $\mu \in [2.03, 5.93]$, a range which includes the MSE=0.15 of regression-estimated $\mu_{SGH}=6.03$ (green colour), meanwhile the MSE corresponding to the arithmetic mean $\mu=8.1$ (red colour) is 0.17. \textbf{Bottom panel:} Estimated $R_0$ and $k$ and line of mean ratio of single to co-colonization. On the left, we see estimated values of $R_0$ and $k$ for all studies and the fitted line of mean $\mu$. On the right, we see a plot of the error for each country, from the SGH model assuming $\mu=E[I/D]=8.1$. Horizontal line with green colour is the Mean Squared Error, and error boundaries to see which countries have the smallest error. In brackets is the percentage of countries with lower values of error, respectively 0.14 and 0.4.}
\label{fig:crudeoptim}
\end{figure*}

\begin{figure*}[!h]
\centering 
\includegraphics[width=0.5\linewidth]{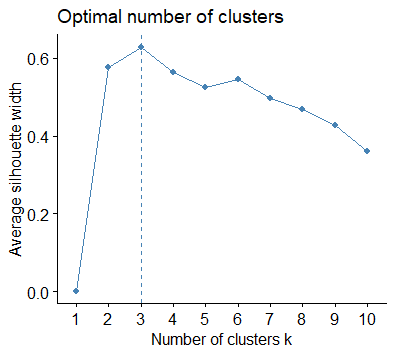}
\caption{Resulting number of clusters by the clustering algorithm in R, applied to 20 studies, defined by their empirical ratios of single to co-colonization $I/D$. Even though the optimal number of clusters appears to be 3, we have opted for favouring a more coarse clustering in 2 groups of studies, where there emerge to be a more balanced number of studies in each cluster, aiding significance of the linear relationship between $\log(k)$ and $\log(R_0-1)$ and hence increasing statistical power. However to see results from applying SGH to 3 clusters of studies separately see Figure \ref{fig:3clustSGH}.}
\label{fig:optimalclusters}
\end{figure*}

\begin{figure*}[!h]
\centering 
\includegraphics[width=0.8\linewidth]{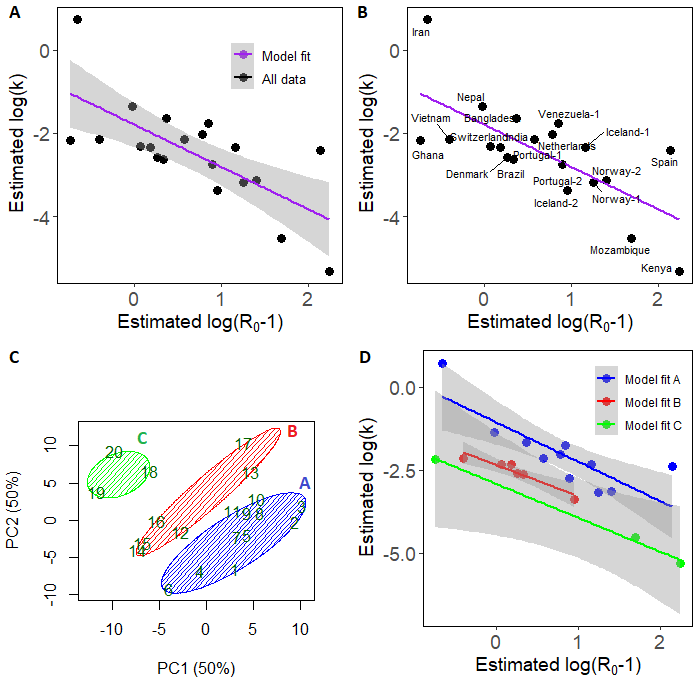}
\caption{\textbf{Analysis and model application with 3 clusters of studies according to $\mu=I/D$ values.} From the clustering algorithm, when studies are clustered in 3 clusters, we obtain the following division of countries. In group A, we have: "Iran", "Spain", "Venezuela-1" "Iceland-1", "Netherlands" "Bangladesh", "Nepal", "Portugal-1", "Norway-2"    "Portugal-2", "Norway-1"), with $\mu_{SGH}^A$ equal 2.89 (p-value=0.00578) and slope equal to -1.1831 (p-value= 0.00208). In Group B, we have: "India", "Switzerland", "Brazil", "Denmark", "Iceland-2", "Vietnam", with $\mu_{SGH}^B$ equal to 10.49 (p-value=6.08e-06) and slope equal to -0.93770 (p-value=0.00434).  
In group C, we have only 3 countries: "Mozambique" "Ghana" "Kenya", with $\mu_{SGH}^C$ equal to 18.41, (p-value=0.0268) and slope equal to -1.02001, (p-value=0.0455). 
When fitting a single global model with cluster id. as an additional variable, results of $lm(\log(k)\sim \log(R_0-1) \times cluster.id)$ point to a model where cluster-id is a significant variable, affecting the intercept but not the slope (no interaction): p-value for intercept is equal to  6.39e-06, for $\log(R_0-1)$ the p-value is 0.0449, for clustering p-value=0.0101. (F-statistic=15.82,p-value= 4.802e-05; Multiple R-square: 0.7479,Adjusted R-square: 0.7006.) Hence, a model with 3 clusters has a smaller $R^2$ finally than a model with 2 clusters, which we favor and present in the main text. The model with 2 clusters enables also more statistical power to explore patterns within two groups of studies of more balanced sizes. }
\label{fig:3clustSGH}
\end{figure*}

\begin{figure*}[!h]
\centering
\includegraphics[width=0.7\linewidth]{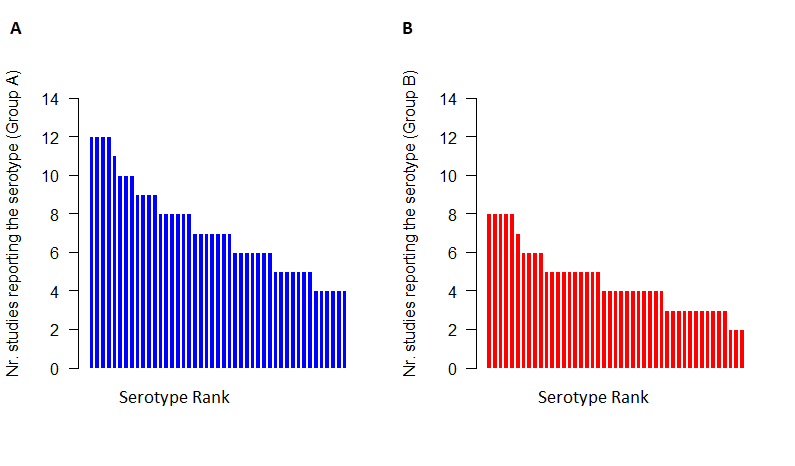}
\caption{\textbf{Empirical rank-frequency distribution for all serotypes reported, by two $\mu$-subgroups.} \textbf{A.} Serotypes occurring in studies of group A (low $\mu$). \textbf{B.} Serotypes occurring in studies of group B (higher $\mu$) - height of the bar indicates in how many studies each serotype appears.}
\label{fig:rankfreq++}
\end{figure*}

\begin{figure*}[!h]
\centering 
\includegraphics[width=0.7\linewidth]{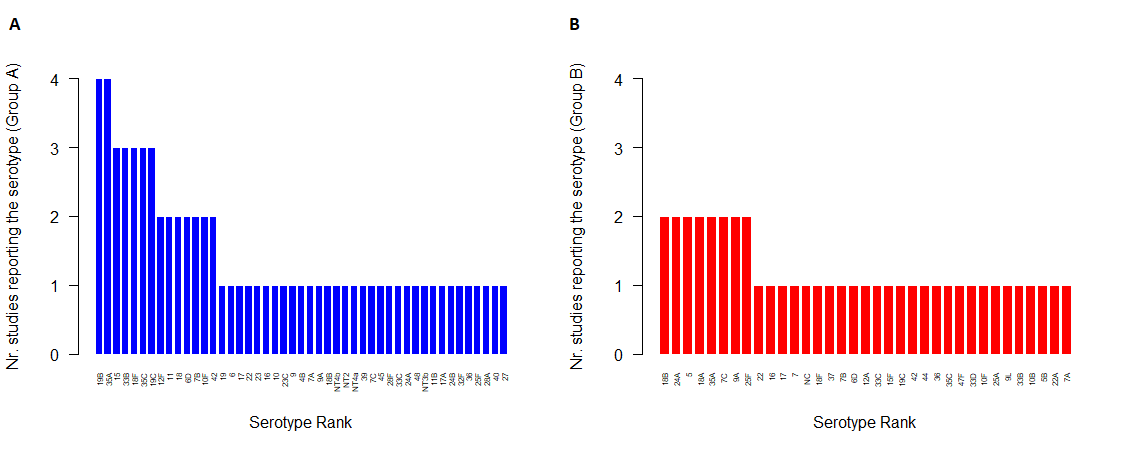}
\caption{\textbf{Identities of rare serotypes in each group}. Serotypes that were reported in $\leq 3$ studies in group A and B.}
\label{fig:lessfrequentserotypes}
\end{figure*}

\begin{figure*}[!h]
\centering 
\includegraphics[width=0.7\linewidth]{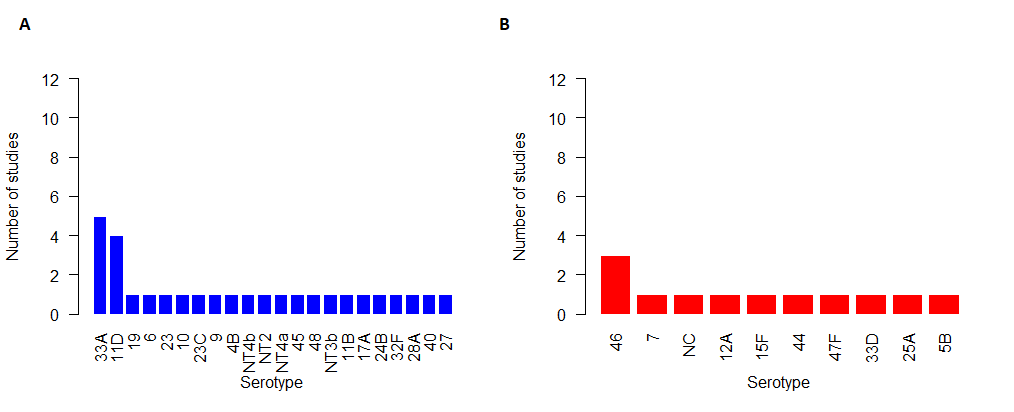} 
\caption{\textbf{Number of occurrences and identities of unique serotypes found exclusively in group A and group B. }}
\label{fig:uniqueserotypesfrequency}
\end{figure*}

\begin{figure*}[!h]
\centering
\includegraphics[width=0.5\linewidth]{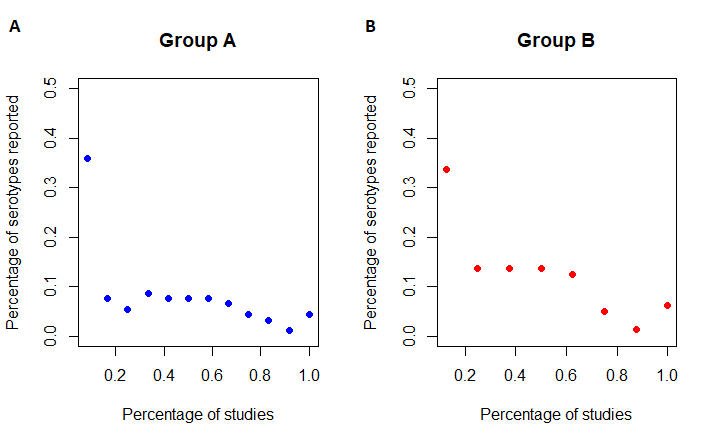}
\caption{\textbf{What proportion of serotypes are found in what proportion of studies?} We plot in percentage form, the global occurrence patterns of different serotypes in the two groups of studies A (low $\mu$) and B (higher $\mu$). For each serotype 92 vs. 80 in groups A and B respectively, we compute in how many studies it is reported (a maximum of 12 vs. 8) in groups A and B respectively. And then we compute statistics over the proportion of serotypes that appear a given proportion of studies. We notice that in each group, about 30\% of serotypes appear in just 1 study, hence pointing to a very prominent site-specific serotype occurrence pattern.}
\label{fig:propser++}
\end{figure*}

\begin{figure*}[!h]
\centering 
\includegraphics[width=0.75\linewidth]{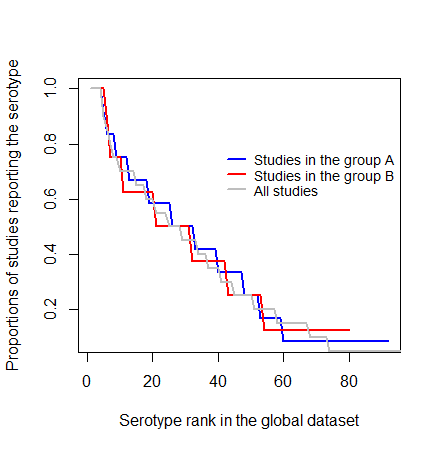}
\caption{\textbf{Serotype distribution according to rank of occurrence by $\mu$ grouping. A: low $\mu$ (blue), B: high $\mu$ (red) and overall (gray).}}
\label{fig:plotSerotypeRank}
\end{figure*}

\begin{table*}[!h]
\caption{\label{tab:Serotypes} Serotypes reported in both groups of studies, with the respective number of times they were reported as present in these groups. Serotypes found only in group A are 22 serotypes: "19"   "6"    "23"   "10"   "23C"  "9"    "4B"   "11D"  "33A"  "NT4b" "NT2"  "NT4a" "45"   "48"   "NT3b" "11B"  "17A"  "24B"  "32F"  "28A"  "40"   "27". Serotypes found only in group B are 10 serotypes: "7"   "NC"  "46"  "12A" "15F" "44"  "47F" "33D" "25A" "5B". The other 70 serotypes are shared among low and high single-to-co-colonization ratio sites. } 
\begin{tabular}{p{1cm}p{7.5cm}p{7.5cm}}

 \hline 
\textit{Nr. of times} & \textbf{Group A (low $\mu=3.4$) -12 sites } & \textbf{Group B (high $\mu=12.4$) - 8 sites }\\
\textit{ } & \textit{Iran, Spain, Venezuela-1, Iceland-1, Netherlands, Bangladesh, Nepal, Portugal-1, Norway-2, Portugal-2, Norway-1, India} & \textit{Switzerland, Brazil, Denmark, Iceland-2, Vietnam, Mozambique, Ghana, Kenya}\\
 \hline
 1&10, 11B, 16, 17, 17A, 18B, 19, 22, 23, 23C, 24A, 24B, 25F, 27, 28A, 28F, 32F, 33C, 36, 39, 40, 45, 48, 4B, 6, 7A, 7C, 9, 9A, NT2, NT3b, NT4a, NT4b&10B, 10F, 12A, 15F, 16, 17, 18F, 19C, 22, 22A, 25A, 33B, 33C, 33D, 35C, 36, 37, 42, 44, 47F, 5B, 6D, 7, 7A, 7B, 9L, NC \\
2&10F, 11, 12F, 18, 42, 6D, 7B&15, 18A, 18B, 22F, 24A, 25F, 29, 35A, 5, 7C, 9A \\
3&15, 18F, 19C, 33B, 35C&1, 11, 13, 19B, 24F, 28F, 31, 33F, 35F, 39, 46 \\
4&10B, 11D, 19B, 22A, 29, 35A, 7F, 9L&10A, 12F, 15A, 18, 21, 23B, 34, 35B, 38, 7F, 8 \\
5&1, 15A, 18A, 20, 24F, 33A, 5&11A, 15B, 15C, 16F, 17F, 18C, 20, 6C, 9N, NT \\
6&13, 17F, 23B, 31, 35F, 37, 38&19A, 23A, 9V \\
7&22F, 23A, 33F, 34, 35B, 8, 9N&3 \\
8&10A, 15C, 16F, 21, 3, 4&14, 19F, 23F, 6A, 6B \\
9&11A, 15B, 18C, 6C& \\
10&23F, 9V, NT& \\
11&19A& \\
12&14, 19F, 6A, 6B& \\
[1ex]
 \hline
\end{tabular}
\end{table*}

\begin{table*}[!h]
\caption{\label{tab:Serogroups} Pneumococcal serogroups reported in group A and group B with the respective number they appeared in these groups. I remove "NT2, NT3b, NT4a, NT4b" from the group A that are the non-typable serotypes reported in Nepal study \cite{kandasamy2015multi}. NC is reported in Denmark study. The number of times serotype 6A is reported will be summed with other number of times other serotypes like 6B are reported and will be considered as the serogroup 6. The same procedure is followed for all other serotypes. See Figure \ref{fig:serogr3} for a visualization. There are 3 serogroups found only in the group B:"46" "44" "47", and 5 serogroups found only in group A are: "45" "48" "32" "40" "27". The other 36 serogroups are shared across low $\mu$ and high $\mu$ settings.} 
\begin{tabular}{p{1cm}p{7.5cm}p{7.5cm}}
 \hline 
\textit{Nr. of times} & \textbf{Group A (low $\mu$) } & \textbf{Group B (high $\mu$) }\\
\textit{ } & \textit{Iran, Spain, Venezuela-1, Iceland-1, Netherlands, Bangladesh, Nepal, Portugal-1, Norway-2, Portugal-2, Norway-1, India} & \textit{Switzerland, Brazil, Denmark, Iceland-2, Vietnam, Mozambique, Ghana, Kenya}\\
\hline
1&25, 27, 28, 32, 36, 39, 40, 45, 48&36, 37, 42, 44, 47  \\
2&12, 42&25, 29, 5  \\
3&&1, 13, 22, 28, 31, 33, 39, 46  \\
4&29&10, 12, 21, 24, 34, 35, 38, 8  \\
5&1, 20, 5&20, 7, NT  \\
6&13, 24, 31, 37, 38, 7&16, 17, 4  \\
7&17, 34, 8&3, 9  \\
8&21, 3, 33, 35, 4&11, 14, 15, 18, 19, 23, 6  \\
9&10, 16, 22&  \\
10&NT&  \\
11&11, 15, 18&  \\
12&14, 19, 23, 6, 9&  \\
[1ex]
 \hline
\end{tabular}
\end{table*}

\begin{figure*}[!h]
\centering 
\includegraphics[width=1.1\linewidth]{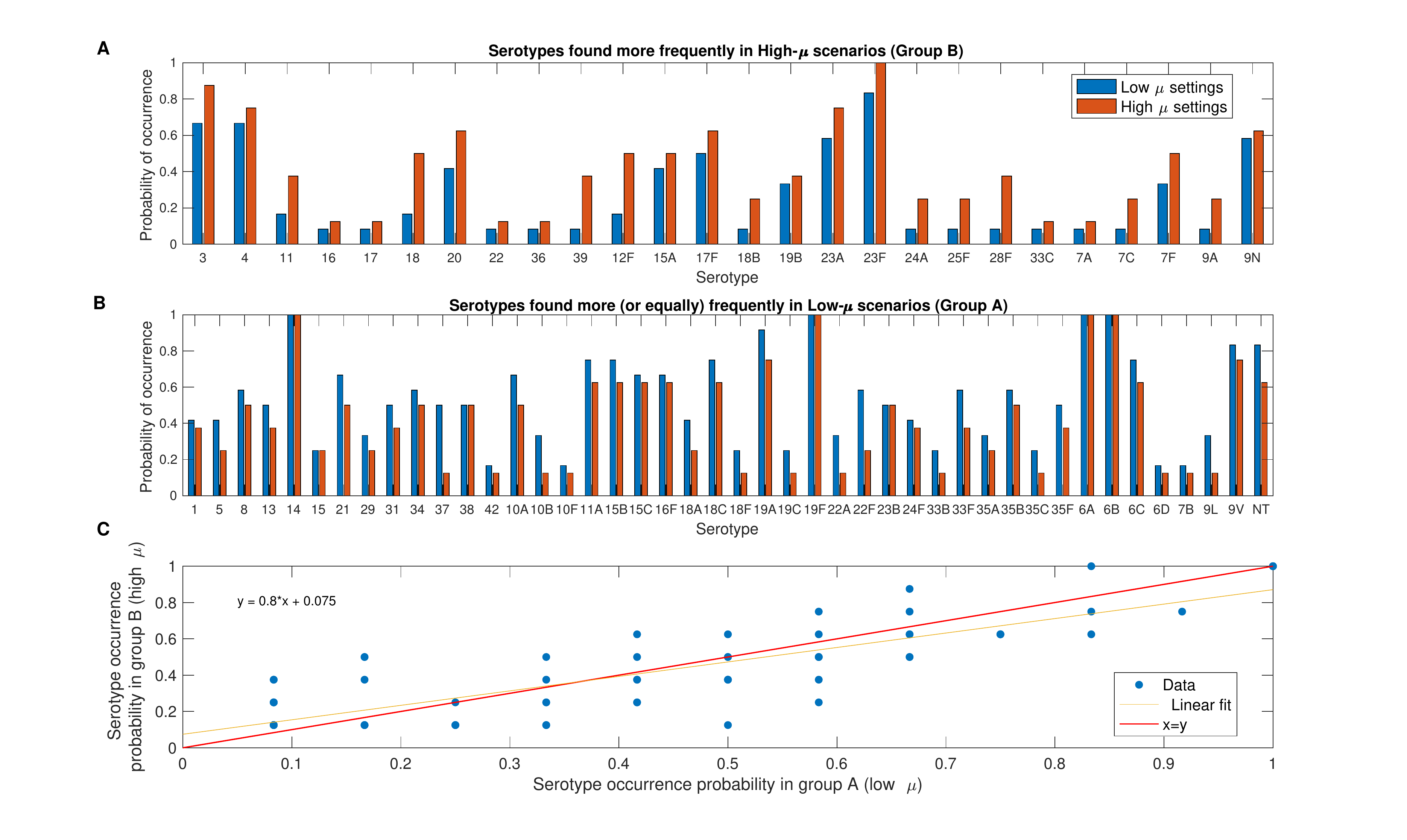} 
\caption{\textbf{Serotypes preferentially occurring in settings of high and low ratio of co-colonization.} \textbf{A.} Here we display serotypes and their frequencies, for those found more frequently in group B studies (high $\mu$: depicted in red). \textbf{B.} Here we display serotypes and their frequencies, for those found more frequently in group A studies (low $\mu$: depicted in blue) \textbf{C.} Here we verify that serotypes that occur more frequently in low $\mu$ studies also occur more frequently in high $\mu$ settings. Despite the patterns in A and B, none of the serotypes showed a significant statistical association with group A or group B, when testing using the Fisher's exact test at 0.05 significance level. Data consistent with Table \ref{tab:Serotypes}.}
\label{fig:lowhighmu1}
\end{figure*}

\begin{figure*}[!h]
\centering 
\includegraphics[width=0.75\linewidth]{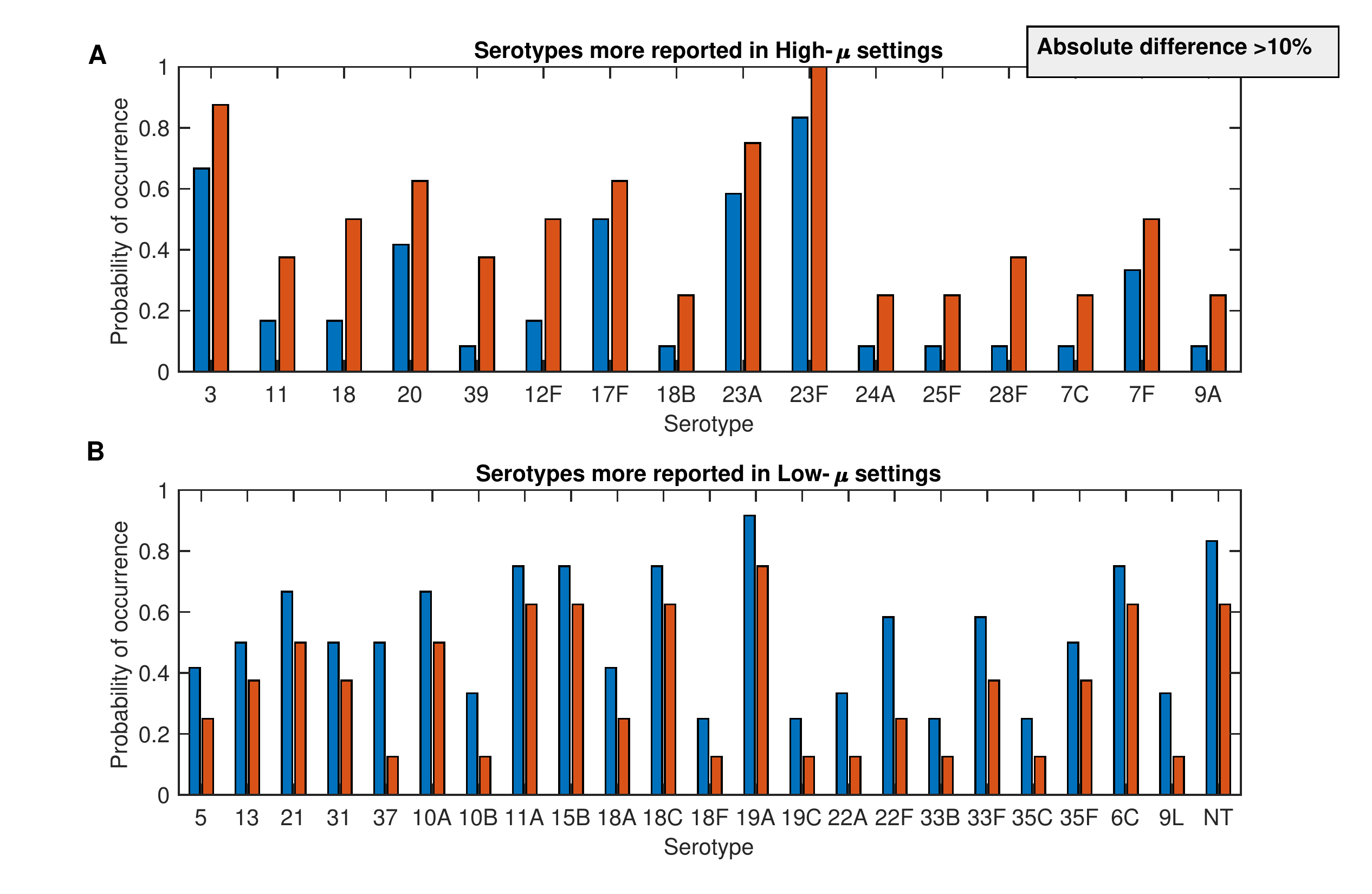} 
\caption{\textbf{Serotypes differing by more than 10\% probability of occurrence in settings of high and low ratio of co-colonization.} \textbf{A.} Here we display serotypes and their frequencies, for those found more frequently in group B studies (high $\mu$: depicted in red). \textbf{B.} Here we display serotypes and their frequencies, for those found more frequently in group A studies (low $\mu$: depicted in blue). Data consistent with Table \ref{tab:Serotypes}.}
\label{fig:lowhighmu2}
\end{figure*}

\begin{figure*}[!h]
\centering 
\includegraphics[width=0.75\linewidth]{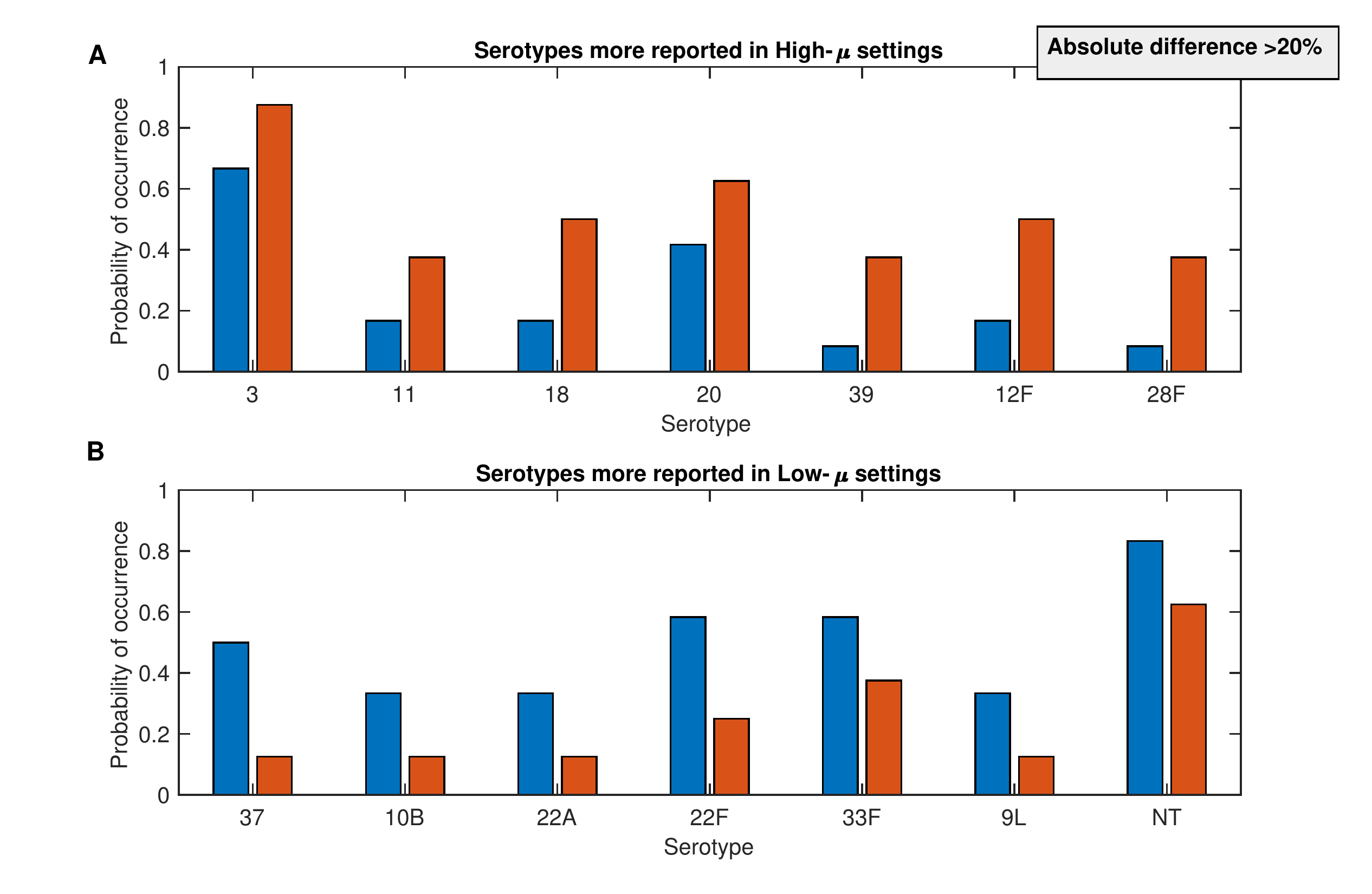} 
\caption{\textbf{Serotypes differing by more than 20\% probability of occurrence in settings of high and low ratio of co-colonization.} \textbf{A.} Here we display serotypes and their frequencies, for those found more frequently in group B studies (high $\mu$: depicted in red). \textbf{B.} Here we display serotypes and their frequencies, for those found more frequently in group A studies (low $\mu$: depicted in blue). Data consistent with Table \ref{tab:Serotypes}.}
\label{fig:lowhighmu3}
\end{figure*}

\begin{figure*}[!h]
\centering 
\includegraphics[width=1\linewidth]{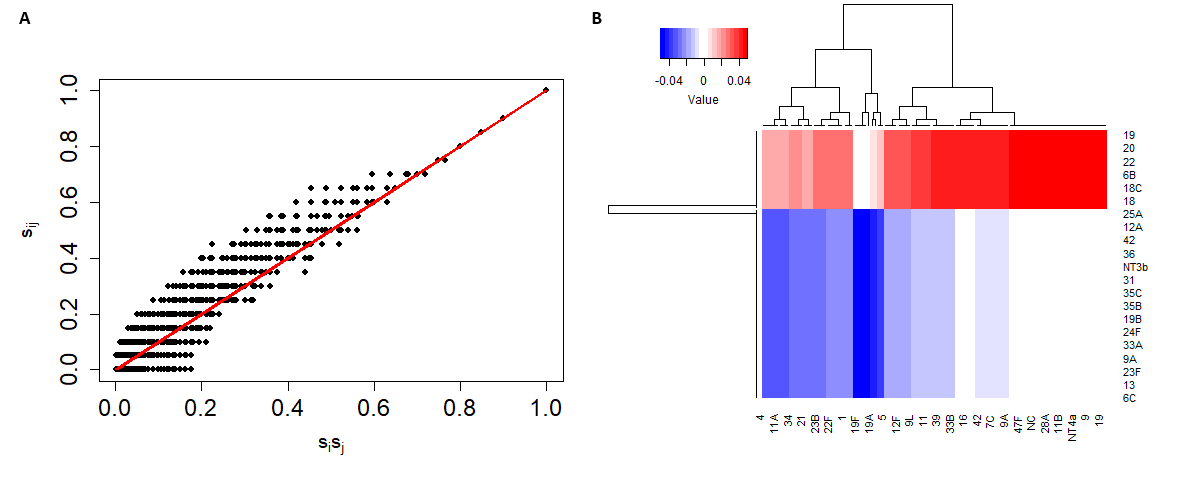} 
\caption{\textbf{Serotype co-occurrence in the same study vs. expectation by chance.} \textbf{A.} Here we display the scatter plot of the product of probabilities of reporting serotype $i$ and $j$ in general in the dataset ($S_i \times S_j$), vs. the probability of co-reporting (co-occurrence of serotypes $i$ and $j$ in the same site $S_{ij}$), while using serotype's occurrence and co-occurrence frequencies across all studies. All the serotypes pairs above the diagonal are found to co-occur together in the same site more often than expected by chance, under the independence assumption, and all the serotype pairs below the diagonal are found to co-occur in the same site, less often than expected by global chance. \textbf{B.} Here we plot the heatmap of serotype pairs co-occurrence distance from independence expectation ($S_{ij}-S_iS_j$), to indicate which serotype pairs belong to which category.}
\label{fig:heatmapandSiSjSij}
\end{figure*}

\begin{figure*}[!h]
\centering
\includegraphics[width=0.7\linewidth]{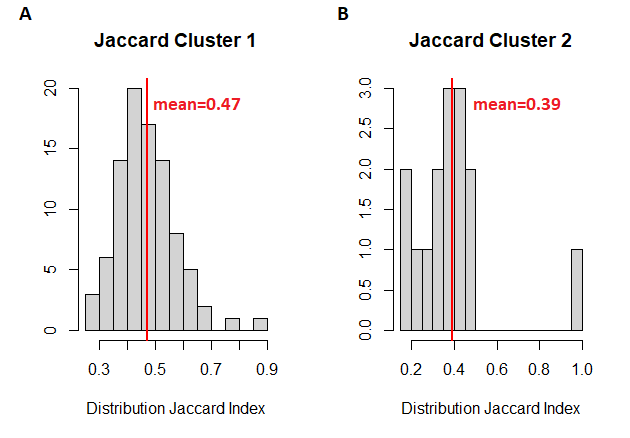}
\caption{\textbf{Clusters of epidemiological studies emerging from Jaccard Index similarity}. Two groups of studies emerged: Group 1: ("Norway-2"    "Norway-1"    "Denmark"     "Portugal-1"  "Portugal-2"  "Mozambique"  "Kenya"       "India"       "Ghana"       "Nepal" "Netherlands" "Spain"       "Bangladesh"  "Brazil" ); and group 2: "Iran"        "Venezuela-1" "Iceland-1"   "Switzerland" "Iceland-2"
 "Vietnam". In A and B we plot the distributions of Jaccard indices within each cluster. Mean Jaccard index in cluster 1 is 0.47 and mean Jaccard index in cluster 2 is 0.39. These clusters do not display significantly different epidemiological characteristics (see Figure \ref{fig:clustJac2}).}
\label{fig:distJac2}
\end{figure*}

\begin{figure*}[!h]
\centering
\includegraphics[width=0.85\linewidth]{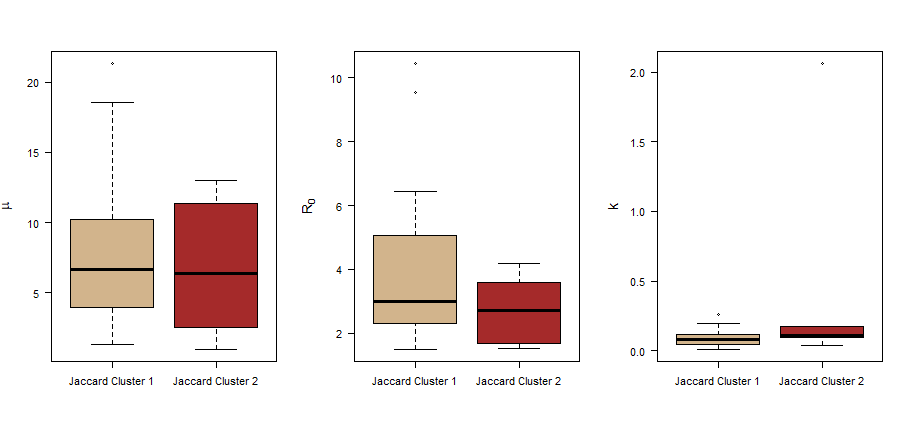}
\caption{\textbf{Studies with different serotype similarity have similar epidemiological parameters.} After obtaining two clusters of studies based on Jaccard index similarity, we tested for differences in single-to-cocolonization ratio $\mu$, basic reproduction number $R_0$ and susceptibility to co-colonization parameter $k$. None of these were significant. Mean single-to-cocolonization ratio $\mu$ in the first group is 8.68, whereas the mean $\mu$ in studies of the second group is 6.77. These were not significant, as a confirmed by a t-test ( p-value=0.4884). Similarly, no significant differences were observed between the basic reproduction numbers or mean susceptibilities to co-colonization between these groups (t-test between $R_0$'s in two groups, p-value=0.1216); and t-test between $k$ coefficients in the two groups, p-value=0.3518).}
\label{fig:clustJac2}
\end{figure*}

 \begin{figure*}[!h]
\centering
\includegraphics[width=0.7\linewidth]{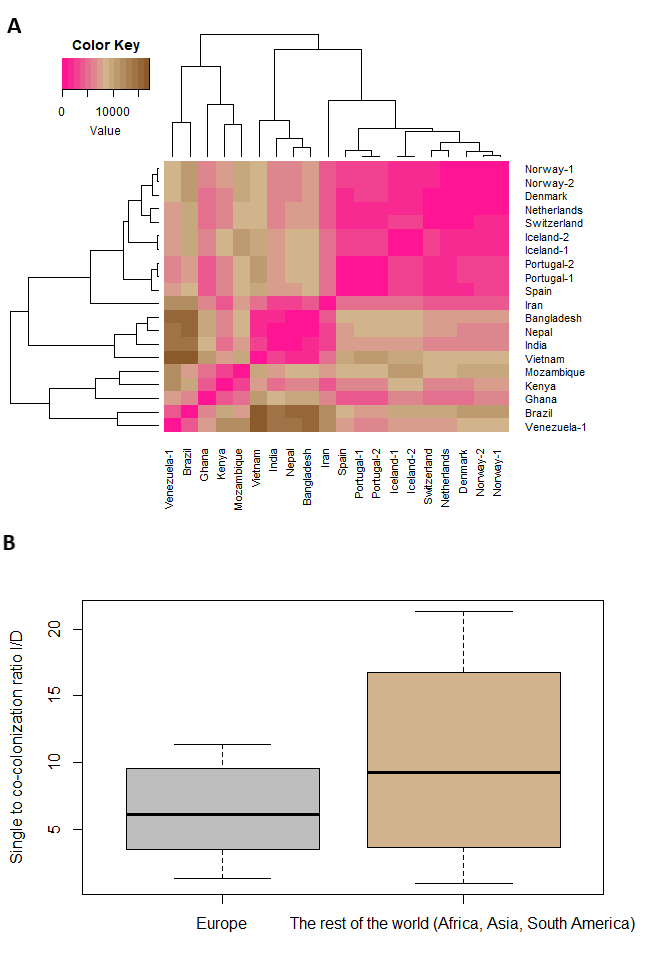}
\caption{\textbf{Geographic distances between study locations and single-to-cocolonization ratios $\mu$}. A. Heatmap of all geographic distances between countries and corresponding dendrogram for hierarchical clustering. B. We plot the distributions of empirical single-to-co-colonization ratios $\mu$ in the two big clusters of studies separated by geographic distance. The two groups were not found to be significantly different by $\mu$, suggesting that geographic distance alone is not a strong determinant of $\mu$ variation. Study locations including Europe and Algeria on average reported a $\mu= 6.3$, study locations including the rest of the world, on average reported a higher value of single to co-colonization $\mu=9.9$.}
\label{fig:geoheatmap}
\end{figure*}

 \begin{figure*}[h]
\centering
\includegraphics[width=0.6\linewidth]{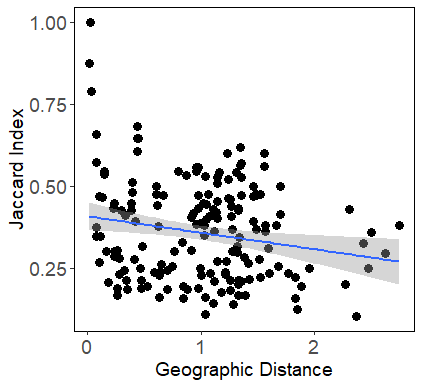}
\caption{\textbf{Geographic distance between studies and Jaccard Index for serotype composition similarity}. We tested whether there is a linear trend between serotype similarity reported across studies and their geographic distance. Applying linear regression in R (Model: $y=a+bx$) we found a significant negative trend ($a= 0.40698$, $b = -0.05014$), with p-value =0.008114. Pearson correlation value between these two variables is -0.1915374 (p-value=0.008). Geographic distance is calculated in kilometers and each value is divided by the mean of all these distances. When performing a similar test for the relationship between geographic distances for any pair of studies and their dissimilarity in epidemiological parameters $R_0$ and $k$ we found no statistically-significant trend. } 
\label{fig:geoscatJ}
\end{figure*}

\begin{figure*}[!h]
\centering 
\includegraphics[width=0.9\linewidth]{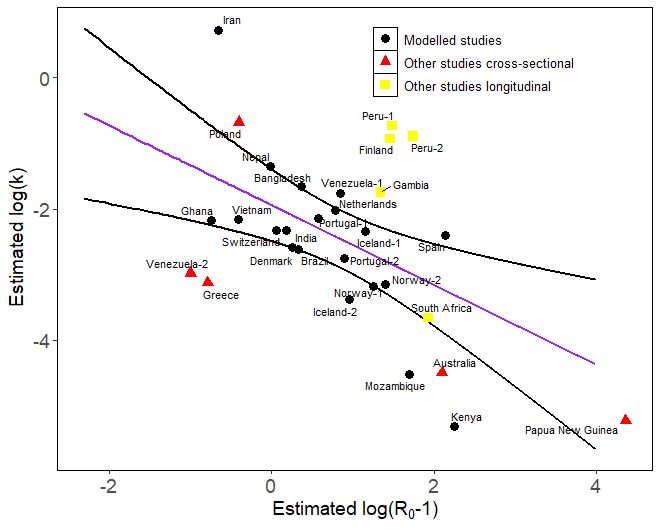} 
\caption{Here we see the prediction from the model $lm(\log(k) \sim \log(R_0-1))$  for $R_0 \in[ 0.1,54]$ with just one global fit for SGH, with model-fitted data (in black), and superimpose all the values corresponding to other studies not included in the model fit (See Table \ref{tab:Otherstudies}): in red, cross-sectional studies in the literature, and in yellow, longitudinal carriage studies of pneumococcus, reporting co-colonization.}
\label{fig:otherpred}
\end{figure*}
\begin{table*}[!h]
\caption{\label{tab:Otherstudies} Other studies that were not included in the model fit. Some of these were excluded based on the type of study (longitudinal (L) vs. crossectional (C)), since we focused on cross-sectional studies to apply our epidemiological transmission model. A few other cross-sectional studies were found later. Two cross-sectional studies were excluded initially based on outlier analysis of the original dataset. See figure \ref{fig:otherpred}, for a plot with these new studies superimposed.} 
\begin{tabular}{p{1.9cm}p{0.9cm}p{1.9cm}p{1.8cm}p{1.75cm}p{0.6cm}p{0.6cm}p{0.6cm}p{0.5cm}p{1.6cm}p{1.1cm}}
 \hline 
\textit{Location} & \textit{Sample Size } & \textit{Susceptible} & \textit{Single-col.}  & \textit{Co-col.}  & \textit{Ratio I/D } & \textit{$R_0$}  & \textit{k}  &\textit{Type} & \textit{Year} & \textit{Ref.}  \\

\textit{ } & \textit{n} & \textit{$n_S(S)$} & \textit{$n_I(I)$} & \textit{$n_D(D)$} & \textit{$\mu$} & \textit {}  & \textit{} & \textit{}  & \textit{}  & \textit{}  \\
\hline
Venezuela-2&1004&733 (73\%)&266 (26.5\%)&5 (0.5\%)&53.2&1.37&0.051&C.  & 2006-2008 & \cite{rivera2011carriage} \\
Greece&2448&1682 (68.7\%)&751 (30.7\%)&15 (0.6\%)&50.07 &1.45&0.044&C.  & 1997-1999 & \cite{syrogiannopoulos2002antimicrobial} \\
Poland&394&236 (59.9\%)&118 (29.9\%)&40 (10.2\%)&2.95&1.67&0.506&C.  & 2016-2020 & \cite{wrobel2022pneumococcal} \\
Australia&174&19 (10.9\%)&142 (81.6\%)&13 (7.5\%)&10.9 &9.1 &0.011&C. & 1981 &  \cite{hansman1985pneumococcal} \\
P.N Guinea &158&2 (1.3\%)&110 (69.6\%)&46 (29.1\%)&2.39&79&0.005&C. & 1985-1987 &\cite{gratten1989multiple,montgomery1990bacterial}\\
Finland&329&62 (18.8\%)&99 (30.1\%)&168 (51.1\%)&0.59&5.30&0.394&L. & 1994-1995 & \cite{syrjanen2001nasopharyngeal} \\
S. Africa&174&22 (12.6\%)&129 (74.1\%)&23 (13.2\%)&5.61&7.91&0.026&L. & 2012-2013 &  \cite{manenzhe2020characterization} \\
Gambia&1553&321 (20.7\%)&739 (47.6\%)&493 (31.7\%)&1.50&4.84&0.174&L. & 2008-2009 &  \cite{chaguza2021carriage} \\
Peru-1&509&94 (18.5\%)&133 (26.1\%)&282 (55.4\%)&0.47&5.41&0.48&L. & 2011 &  \cite{saha2015detection} \\
Peru-2&506&75 (14.8\%)&129 (25.5\%)&302 (59.7\%)&0.43&6.75&0.407&L.  & 2009 &  \cite{saha2015detection}  \\
 \hline
\end{tabular}
\end{table*}

\end{document}